\pdfoutput=1
\documentclass[iop,twocolumn,twocolappendix,numberedappendix]{openjournal}
\usepackage{needspace}  
\usepackage{graphicx}
\usepackage{listings}
\usepackage{placeins}
\usepackage{float}
\usepackage{booktabs}
\usepackage{lineno}
\usepackage{amsmath,amstext}
\usepackage[T1]{fontenc}
\usepackage[utf8]{inputenc}
\usepackage[flushleft]{threeparttable}
\usepackage{url}
\usepackage{xspace}
\usepackage[colorlinks, citecolor=blue]{hyperref}

\usepackage[figure,figure*]{hypcap}

\usepackage{lsstdesc_macros}
\usepackage{graphicx}
\usepackage{hyperref}
\usepackage{url}
\usepackage{natbib}
\usepackage{tikz}
\usepackage{amssymb}
\usepackage{cases}
\usepackage[normalem]{ulem}
\usepackage{xcolor}
\graphicspath{{./}{./figures/}}

\usepackage{comment}
\usepackage{multirow}

\usetikzlibrary{fit}
\usetikzlibrary{backgrounds}
\usetikzlibrary{arrows.meta}
\usetikzlibrary{positioning}
\definecolor{PipelineColor}{HTML}{2563EB}       
\definecolor{PipelineColorfill}{HTML}{DBEAFE}   
\definecolor{DataProductColor}{HTML}{9333EA}    
\definecolor{DataProductColorfill}{HTML}{F3E8FF}
\definecolor{DataProductOutColor}{HTML}{7f6000} 
\definecolor{importGray}{HTML}{9CA3AF}          
\definecolor{provenanceGray}{HTML}{4B5563}      
\def\dfsize{\sffamily}
\newcommand{\diagText}[1]{\texttt{#1}}
\newcommand{\diagTitle}[3][black]{ \scalebox{1.4}{ \textbf{ \color{#1}{#2} } } }
\tikzset{%
    >={Latex[width=2.5mm,length=2mm]},
    pipe/.style = {
        draw=PipelineColor,
        fill=PipelineColorfill,
        minimum width=3cm, minimum height=1cm,
        text width=3.8cm,
        text centered,
        line width=1.5pt,
        font=\sffamily\small},
    pipeline/.style = {rectangle, pipe},
    pipelineII/.style = {rectangle, pipeline, minimum height=1.4cm},
    pipelineMod/.style = {rectangle, pipeline, dashed},
    dataproduct/.style = {
        rounded corners,
        pipe,
        minimum width=.5cm,
        text width=3.cm,
        draw=DataProductColor,
        fill=DataProductColorfill,
        },
}
\newcommand{\Firecrown}[1]{
    \node(Firecrown)[#1]{
        \diagTitle{Firecrown}\\
        (Construct Likelihood)};
}
\newcommand{\TJPCov}[1]{
    \node(TJPCov)[#1]{
        \diagTitle{TJPCov}{}\\
        \diagText{(Theoretical covariance)}};
}
\newcommand{\Theory}[1]{
    \node(Theory)[#1]{
        \diagTitle[black]{CROW}{}\\
        \diagText{(HMF, MoR, SF)\\
        $\langle N\rangle_i$ +
        {\color{blue} $\langle M\rangle_i$}/%
        {\color{black} $\langle \Delta\Sigma\rangle_i$}/%
        {\color{DataProductOutColor} $\langle g_t\rangle_i$}}
        };
}
\newcommand{\CCL}[1]{
    \node(CCL)[#1]{
        \diagTitle[black]{CCL}{}\\
        \diagText{(Theoretical Cosmology)}
};}
\newcommand{\CLMM}[1]{
    \node(CLMM)[#1]{
        \diagTitle{CLMM}{}\\
        \diagText{(Cluster WL Masses)}};
}
\newcommand{\TXPipeDataWL}[1]{
    \node(TXPipeDataWL)[#1]{
        $(e_1, e_2)$ /
        $\Delta\Sigma$, $g_t$
        };
}
\newcommand{\TXPipeDataNC}[1]{
    \node(TXPipeDataNC)[#1]{
        $\langle N\rangle_i$ };
}
\newcommand{\CFData}[1]{
    \node(CFData)[#1]{
        Cluster Catalog};
}
\newcommand{\InDataDM}[1]{
    \node(InDataDM)[#1]{
        Galaxy Catalog};
}
\def\arrwII{-{Latex[length=3mm, width=3mm]}}
\newcommand{\LINEimports}[2]{\draw[line width=2pt, dashed, color=importGray, \arrwII](#1) -- (#2);}
\newcommand{\LINEprovenance}[2]{\draw[line width=1pt, color=provenanceGray, \arrwII](#1) -- (#2);}
\newcommand{\LINEdataproducts}[2]{\draw[line width=1pt, color=black, -](#1) -- (#2);}
\newcommand{\TXPipeNewer}{
    \node(TXPipeEnsambleMod)[pipelineMod, above=0.5 of CLMMData]{
        \diagTitle{CL Ensemble}\\
    };
    \TXPipeDataWL{dataproduct, above left=0.3 and -1.5 of TXPipeEnsambleMod, text width=2cm}
    \TXPipeDataNC{dataproduct, above right=0.3 and -1.5 of TXPipeEnsambleMod, text width=1cm}
    \node(TXPipeDataMod)[pipelineMod, above=1.5 of TXPipeEnsambleMod]{\diagTitle{Data Retrieval}\\};
    \node(TXname)[pipeline, above=0.1 of TXPipeDataMod, draw=PipelineColorfill]{
        \diagTitle{TXPipe} ((Data measurements)
    };
    \begin{pgfonlayer}{background}
        \node(TXPipe)[pipeline, fit=(TXname) (TXPipeEnsambleMod), inner sep=6pt, text width=6cm]{};
    \end{pgfonlayer}
  \LINEprovenance{TXPipeDataWL}{TXPipeEnsambleMod}
  \LINEprovenance{TXPipeDataNC}{TXPipeEnsambleMod}
  \LINEdataproducts{TXPipeDataMod}{TXPipeDataWL}
  \LINEdataproducts{TXPipeDataMod}{TXPipeDataNC}
}
\def\nodewidthPipe{3cm}
\newcommand{\diagramGrenoble}{
    \begin{tikzpicture}[node distance=1cm,
        every node/.style={fill=white, font=\dfsize}, align=center]
    \Firecrown{pipeline, text width=4cm}
    \Theory{pipelineII,text width=\nodewidthPipe, above right=1.94 and -2 of Firecrown}
    \TJPCov{pipelineII,text width=\nodewidthPipe, right=1 of Theory}
    \CLMM{pipelineII,text width=\nodewidthPipe, above=1 of Theory}
    \CCL{pipelineII,text width=\nodewidthPipe, above=1 of TJPCov, right=1 of CLMM}
    \node(CLMMData)[dataproduct,text width=3cm, above left=-0.5 and 1 of Firecrown]{
        $\langle N\rangle_i$ + \\
        {\color{blue} $\langle M\rangle_i$}/%
        {\color{black} $\langle \Delta\Sigma\rangle_i$}/%
        {\color{DataProductOutColor} $\langle g_t\rangle_i$}
        };
    \TXPipeNewer
    \CFData{dataproduct,text width=2.4cm, minimum height=0.7cm, above right=0.5 and -2.5 of TXPipe}
    \InDataDM{dataproduct,text width=2.4cm, minimum height=0.7cm, above left=.5 and -2.5 of TXPipe}
    \LINEimports{CLMM}{Theory}
    \LINEimports{Theory}{Firecrown}
    \LINEimports{CCL}{Theory}
    \LINEimports{CCL}{TJPCov}
    \LINEimports{CCL}{CLMM}
    \LINEimports{Theory}{TJPCov}
    \LINEimports{CLMM}{TXPipeEnsambleMod}
    \LINEimports{TJPCov.south}{Firecrown}
    \LINEprovenance{CLMMData}{Firecrown}
    \LINEprovenance{CFData}{TXPipe}
    \LINEprovenance{InDataDM}{TXPipe}
    \LINEdataproducts{TXPipeDataMod}{TXPipeDataWL}
    \LINEdataproducts{TXPipeDataMod}{TXPipeDataNC}
    \LINEdataproducts{TXPipe}{CLMMData}
    \end{tikzpicture}
}

\newcommand{\clmm}{\code{CLMM}}

\newcommand{\ccl}{\code{CCL}}

\definecolor{strongteal}{HTML}{007F86}

\usepackage{xspace}
\newcommand{\collab}[1]{\textsc{#1}}

\newcommand{\desc}{\collab{LSST-DESC}\xspace}
\newcommand{\pipeline}{\collab{LSST-DESC} Cluster Pipeline\xspace}

\lstdefinelanguage{yaml}{
  keywords={true,True,false,False,null,Null,yes,no},
  keywordstyle=\color{blue}\bfseries,
  basicstyle=\ttfamily\normalsize,
  sensitive=false,
  comment=[l]{\#},
  morecomment=[s]{/*}{*/},
  commentstyle=\color{purple}\ttfamily,
  stringstyle=\color{red}\ttfamily,
  morestring=[b]',
  morestring=[b]",
  breaklines=true,
  showstringspaces=false,
}

\newcommand{\redmapper}{\texttt{redMaPPer}\xspace}

\newcommand\redout{\bgroup\markoverwith
{\textcolor{red}{\rule[.5ex]{2pt}{0.4pt}}}\ULon}

\begin{document}
\title{
Towards the LSST DESC pipeline for cosmology with galaxy clusters -- \\ design and validation on simulations\vspace{-1cm}}


\author{
E.~J.~Barroso$^{1}$,
C.~Payerne$^{2}$,
M.~Aguena$^{3}$,          
C.~Combet$^{4}$,
S.~Elles$^{1}$,
T.~Guillemin$^{1}$,
M.~Ricci$^{5}$,
S.~D.~P.~Vitenti$^{6}$,
J.~Zuntz$^{7}$,
N.~Amouroux$^{1}$,
M.~April$^{8}$,
C.~Avestruz$^{9}$,
D.~Boutigny$^{1}$,
N.~E.~Chisari$^{10,11}$,
A.~Fumagalli$^{3,12}$,
C.~Garc\'ia-Garc\'ia$^{13}$,
N.~Lorenzo~Martinez$^{1}$,
A.~I.~Malz$^{14}$,
M.~Paterno$^{15}$,
E.~M.~Pedersen$^{16}$,
M.~Penna-Lima$^{17,18}$,
E.~S.~Rykoff$^{19,20}$,
N.~\v{S}ar\v{c}evi\'c$^{21}$,
R.~Solomon$^{12}$,
K.~Tsiane$^{9}$,
Y.~Zhang$^{22}$,
and the LSST Dark Energy Science Collaboration
}

\affiliation{
$^{1}$LAPP, Univ. Savoie Mont Blanc, CNRS, Annecy, France\\
$^{2}$IRFU, CEA, Universit\'e Paris-Saclay, F-91191 Gif-sur-Yvette, France\\
$^{3}$INAF -- Osservatorio Astronomico di Trieste, via G. B. Tiepolo 11, I-34143 Trieste, Italy\\
$^{4}$Univ. Grenoble Alpes, CNRS, LPSC-IN2P3, 38000 Grenoble, France\\
$^{5}$Universit\'e Paris Cit\'e, CNRS, Astroparticule et Cosmologie, F-75013 Paris, France\\
$^{6}$Departamento de F\'isica, Universidade Estadual de Londrina, Rod. Celso Garcia Cid, Km 380, 86057-970, Londrina, Brazil\\
$^{7}$Institute for Astronomy, University of Edinburgh, Royal Observatory, Blackford Hill, Edinburgh EH9 3HJ, UK\\
$^{8}$University of California, Santa Cruz, 1156 High Street, Santa Cruz, CA 95064, USA\\
$^{9}$University of Michigan, 500 S State St, Ann Arbor, MI 48109, USA\\
$^{10}$Institute for Theoretical Physics, Utrecht University, Princetonplein 5, 3584 CC, Utrecht, the Netherlands\\
$^{11}$Leiden Observatory, Leiden University, P.O. Box 9513, 2300 RA Leiden, the Netherlands\\
$^{12}$Ludwig-Maximilians-Universit\"at M\"unchen (LMU), Munich, Germany\\
$^{13}$CIEMAT, Avenida Complutense 40, E-28040 Madrid, Spain\\
$^{14}$Space Telescope Science Institute, 3700 San Martin Drive, Baltimore, MD 21218, USA\\
$^{15}$Fermi National Accelerator Laboratory, Batavia, IL 60510, USA\\
$^{16}$Department of Physics, Harvard University, 17 Oxford St., Cambridge, MA 02138, USA\\
$^{17}$Instituto de F\'isica e Centro Internacional de F\'isica, Universidade de Bras\'ilia, 70297-400, Bras\'ilia, Brazil\\
$^{18}$Programa de P\'os-Gradua\c{c}\~ao em F\'isica, Universidade Estadual de Londrina, Rod. Celso Garcia Cid, Km 380, 86057-970, Londrina, Brazil\\
$^{19}$Kavli Institute for Particle Astrophysics \& Cosmology, P.O. Box 2450, Stanford University, Stanford, CA 94305, USA\\
$^{20}$SLAC National Accelerator Laboratory, Menlo Park, CA 94025, USA\\
$^{21}$Department of Physics, Duke University, Science Dr, Durham, NC 27710, USA\\
$^{22}$NSF NOIRLab, 950 N Cherry Avenue, Tucson, AZ 85719, USA\\
Contact: \href{mailto:barroso@lapp.in2p3.fr}{barroso@lapp.in2p3.fr}
}


\begin{abstract}
We present the current implementation of an automated framework developed within the Legacy Survey of Space and Time Dark Energy Science Collaboration (LSST DESC) for galaxy cluster cosmology, in preparation for the first LSST data releases of the NSF--DOE Vera C. Rubin Observatory. The cluster cosmology pipeline integrates DESC-developed tools dedicated to galaxy/cluster catalog ingestion, cluster number count and weak lensing measurements, modeling, and parameter inference through posterior sampling, enabling reproducible end-to-end cluster cosmology analyses. We test and validate the pipeline using synthetic mock datasets and the LSST DESC cosmoDC2 simulation, performing a joint analysis of \redmapper cluster abundance and stacked weak lensing profiles measured in bins of redshift and richness. We constrain two cosmological parameters, $\Omega_c$ and $\sigma_8$, along with six cluster mass--observable relation parameters. Our baseline analysis recovers the fiducial cosmoDC2 values of $\Omega_c$ and $\sigma_8$ within $0.51\sigma$ and $0.34\sigma$, respectively. We also showcase the flexibility of the overall pipeline by assessing the impact of fiducial cosmology and alternative modeling prescriptions on the parameter constraints. Lastly, we also discuss the current limitations of the pipeline and the developments needed for LSST data.

\end{abstract}



\maketitle
\section{Introduction}
\label{sec:intro}

Galaxy clusters are hosted by the most massive dark matter halos, which form through the gravitational collapse of large matter overdensities and are the most massive gravitationally bound structures in the Universe~\citep{kravtsov2012formation}. Their mass function and spatial distribution are sensitive to both the expansion history and the growth of large-scale structure over cosmic timescales, making galaxy clusters a powerful cosmological probe complementary to other probes such as the cosmic microwave background, Type Ia supernovae, baryon acoustic oscillations, and weak gravitational lensing. This sensitivity enables constraints on cosmological parameters and the nature of dark energy and dark matter in the Universe (see~\citealt{allen2011cosmological, weinberg2013observational} for reviews). The primary cosmological probe using galaxy clusters is their abundance in the Universe, which has been used in cosmological analyses of major surveys including \textit{Planck} \citep{ade2016planck}, SPT \citep{deHaan2016,Bocquet2019}, eROSITA \citep{ghirardini2024srg}, DES \citep{desclustercosmo20,abbott2025dark}, HSC \citep{Chiu2024hsccl}, and KiDS~\citep{ lesci2022amico,Lesci2025kidsCL}.

One of the dominant limiting factors in optical cluster cosmology is the combined uncertainty and scatter of the mass-observable relation (MoR)~\citep{allen2011cosmological}, which relates cluster halo mass to observable proxies, as halo mass cannot be directly measured and must instead be inferred~\citep[e.g.,][]{Murata2018lensing, Payerne2025dc2CL}. As a result, cluster abundance alone provides limited constraints on cosmological parameters, due to strong degeneracies with the scaling-relation parameters. Cluster weak gravitational lensing, in which the projected images of distant galaxies are sheared by the gravitational field of foreground clusters, provides a nearly unbiased mass calibration that helps break this degeneracy~\citep{Murata2019, abbott2025dark}. However, there are other limiting factors to this approach that may bias the scaling relation, such as the correlation between cluster observable and the cluster weak lensing profile (e.g.,~\citealt{zhang2024impact}). Other combinations are also being explored, for instance adding cluster clustering and the galaxy two-point statistics in a joint 6$\times$2pt$+$N analysis~\citep{To2021DEScomb}. The joint analysis of cluster abundance and cluster weak lensing is one of the standard approaches adopted by the community to constrain cosmological parameters~\citep{Mantz2015WtG, Chiu2023eFEDS, Bocquet2024spt, fumagalli2024cosmological, ghirardini2024srg}. Carrying it out in a reliable, reproducible, and automated way requires a dedicated analysis pipeline, modular enough to accommodate evolving approaches, a need that becomes especially pressing in the context of the upcoming Stage-IV surveys.

We are transitioning from the successful Stage-III galaxy surveys such as the Dark Energy Survey (DES,~\citealt{desclustercosmo20,To2021DEScomb,Costanzi2021DESSPTCL, abbott2025dark}), the Hyper Suprime-Cam survey (HSC,~\citealt{Murata2019, Sunayama2023HSClensing, Chiu2024hsccl}) and the Kilo-Degree Survey (KiDS,~\citealt{lesci2022amico}), to a new era of Stage-IV lensing surveys with the Legacy Survey of Space and Time (LSST,~\citealt{lsst2012}), the ESA satellite Euclid~\citep{Euclid2025}, and NASA's Nancy Grace Roman Space Telescope~\citep{bailey2023nancy}. These next-generation surveys will deliver unprecedentedly large galaxy samples, enabling the detection of tens of thousands of clusters and high-precision weak gravitational lensing measurements. This work focuses on LSST, which will map $18,000$ square degrees of the southern sky over $10$ years, allowing measurements of shapes and redshifts of a few billion galaxies up to $z \sim 3$. LSST will permit the detection and mapping of the distribution of $\sim 100\,000$ galaxy clusters, as well as characterizing their dark matter content through lensing~\citep{lsst2012, vedder2021galaxy}. 

The LSST Dark Energy Science Collaboration (LSST-DESC, \citealt{lsst2012}) is preparing for the cosmological analysis of LSST data, and has been developing a suite of software tools covering all stages of the analysis chain, from catalog construction and weak lensing measurement to theoretical modeling and parameter inference. Cluster constraints will eventually be combined with the other DESC probes, which requires the cluster analysis to be built on the same tools and calibrations as the rest of the collaboration, so that systematics and covariances are treated consistently across probes. Combining these tools into a validated, end-to-end framework is one of the key challenges in preparing for LSST Data Release 1 (DR1). The significantly reduced statistical uncertainties of LSST compared to Stage-III surveys place stringent requirements on the control of systematic effects in galaxy cluster cosmology. Key examples include galaxy shape calibration~\citep{metacal, McClintock2019mass}, selection effects~\citep{Aguena2018, Rykoff2014redmapper}, and photometric redshift uncertainties in cluster weak lensing~\citep{Applegate2014, Payerne2025dc2CL}, all of which become more critical in the LSST regime. While these effects will need to be addressed progressively as LSST data arrive, a necessary first step is to establish a validated baseline: a working end-to-end analysis framework that correctly combines all the relevant DESC tools at their current version, spanning data ingestion, data vector generation, theoretical modeling, covariance estimation, and Bayesian inference. Each stage must be benchmarked to provide a reliable baseline for future analyses. 
Within DESC, individual tools have already been presented and validated in dedicated code papers, including \texttt{CCL}~\citep{2019ApJS..242....2C}, \texttt{CLMM}~(\citealt{Aguena2021clmmv1, Aguena2026clmmv2}), \texttt{TXPipe}~\citep{Txpipe}, \texttt{RAIL}~\citep{rail_pipeline} and \texttt{Augur}~\citep{rogozenski2026augur}, but no cluster cosmology analysis validating the DESC tools as an integrated system has yet been carried out. 

In this work, we  introduce~\texttt{CLPipe}\footnote{\url{https://github.com/LSSTDESC/CLPipe}}, the current implementation of the \pipeline, designed to unify these DESC tools into a single automated and reproducible analysis chain, in which individual components can be swapped or extended as the analysis evolves, minimizing human error and ensuring reproducibility. We demonstrate its capability to perform a cosmological analysis, from catalog-level data to parameter inference using the simulated cosmoDC2 extragalactic catalog~\citep{korytov2019cosmodc2, lsst2021lsst}. To this end, we analyze the abundance and lensing of \texttt{redMaPPer}-detected clusters in cosmoDC2~\citep{Rykoff2014redmapper}. This work constitutes the first end-to-end cluster cosmology analysis performed entirely within the \pipeline framework, establishing a validated baseline in preparation for the first cosmological data release of LSST.

This paper is organized as follows. Sec.~\ref{sec:clc} presents the ingredients needed for a counts and weak lensing cluster analysis. Sec.~\ref{sec:interface} presents each stage of the \pipeline (connectivity to other DESC tools and performance). We validate the pipeline inference framework with mocks in Sec.~\ref{sec:mock_analysis}. Sec.~\ref{sec:results} presents the analysis of the cosmoDC2 simulated dataset that is used to test and validate the pipeline. In Sec.~\ref{sec:discussion} we discuss our results and perspectives.

\begin{figure*}[tbp]
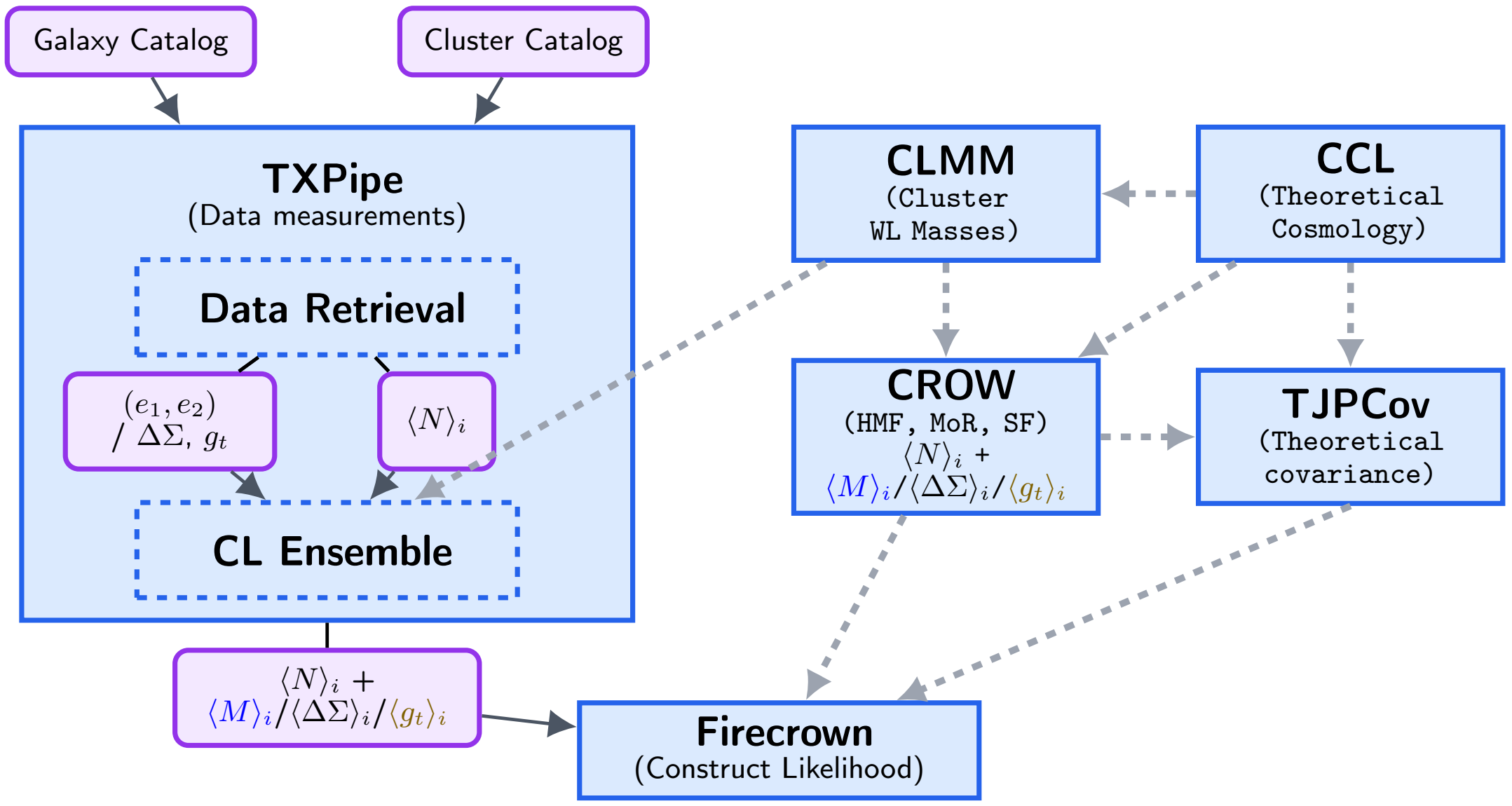

    \centering
    \resizebox{0.8\textwidth}{!}{\diagramGrenoble}
    \caption{Diagram showing the structure of the DESC Cluster Cosmology pipeline. Purple rectangles represent input and output data products, while blue rectangles denote software modules. Solid arrows indicate data passed as inputs, while dashed arrows represent imported modules.}
    \label{fig:cluster_pipeline_diagram}
\end{figure*}

\section{Cosmology with galaxy clusters}
\label{sec:clc}

Extracting cosmological constraints from a cluster catalog requires connecting theory predictions to the estimated data vectors. In this work, we consider the combination of cluster abundance and cluster weak lensing. We describe the general ingredients and steps of this joint analysis below, providing the background needed to understand the construction of the pipeline presented in Sec.~\ref{sec:interface}.

The analysis starts with a cluster catalog containing a set of properties for each cluster, in particular its redshift and an observable that serves as a mass proxy. The abundance is then the distribution of the detected clusters over this mass proxy--redshift plane. Its prediction depends on the survey volume and halo mass function (e.g.,~\citealt{Tinker2008, despali2016universality}), with the mass--observable relation connecting the theoretical halo mass to the observed mass proxy (e.g.,~\citealt{giodini2013scaling}). This prediction may also account for the selection function, describing the incomplete and impure detections of clusters~(e.g.,~\citealt{Aguena2018}), and for other systematics such as projection effects~(e.g.,~\citealt{Costanzi2019, sunayama20}) and cluster photometric redshift uncertainties.

We then estimate the cluster weak lensing data vectors, the cluster shear profiles, which provide a complementary probe to cluster counts. Galaxy clusters act as lenses for the galaxies behind them, distorting their observed shapes, such that the measured average tangential shear of background galaxies around clusters provides information on their underlying mass distribution~\citep{Schneider2005, Bartelmann2001}. Individual cluster shear profiles may be dominated by noise from intrinsic ellipticity of source galaxies (shape noise)~\citep{hoekstra03, Wu2019covDeltaSigma}, and thus they can be stacked over samples of clusters to increase the signal-to-noise ratio~\citep{Mandelbaum2005}. Modeling such profiles requires a description of the matter density around halos~\citep{nfw, Aguena2021clmmv1}, which can include systematic effects such as cluster miscentering~\citep{zhang2023modelling} and the contamination of the source sample by cluster member galaxies~\citep{Applegate2014}.

Once the abundance and lensing data vectors are built, we extract cosmological parameters jointly with the mass--proxy relation parameters and any additional analysis nuisance parameters. This is usually done through likelihood-based inference, where the posterior distribution of the parameters is sampled with Markov Chain Monte Carlo (MCMC) methods~\citep{abbott2025dark, artis2022galaxy, ghirardini2024srg}. The form of the data vectors depends on the approach adopted. In the binned approach, clusters are grouped into fixed intervals of mass proxy and redshift: the abundance data vector is the number of clusters in each interval, and the lensing data vector a stacked profile per interval~\citep{desclustercosmo20, lesci2022amico}. In the unbinned approach, the data vector instead retains the mass proxy and redshift of each detected cluster, with the lensing signal entering at the level of individual cluster profiles~\citep{cash1979parameter, hu2006likelihood, Bocquet2019}. This avoids the information loss from binning and stacking, at the cost of computationally expensive likelihood and covariance evaluations \citep{Payerne2024} for the large catalogs expected from optical surveys.

Finally, cluster abundance and weak lensing are combined by multiplying their individual likelihoods to extract their cosmological information. Each likelihood requires a covariance matrix. For the abundance, the scatter arises from the Poisson noise and from the super-sample covariance~\citep{Hu2003, Lacasa2018, Payerne2023}. For the lensing profiles, the dominant contribution on small scales is the shape noise of the source galaxies~\citep{Wu2019covDeltaSigma}. Both covariances can be computed from analytical models or estimated from the data. As the theoretical covariances depend on the parameters themselves, they can either be held fixed during the inference or recomputed at each step. The \pipeline assembles these analysis steps into a single automated framework, described in the next section.

\begin{figure*}[tbp]
    \centering
    \includegraphics[width=.8\linewidth]{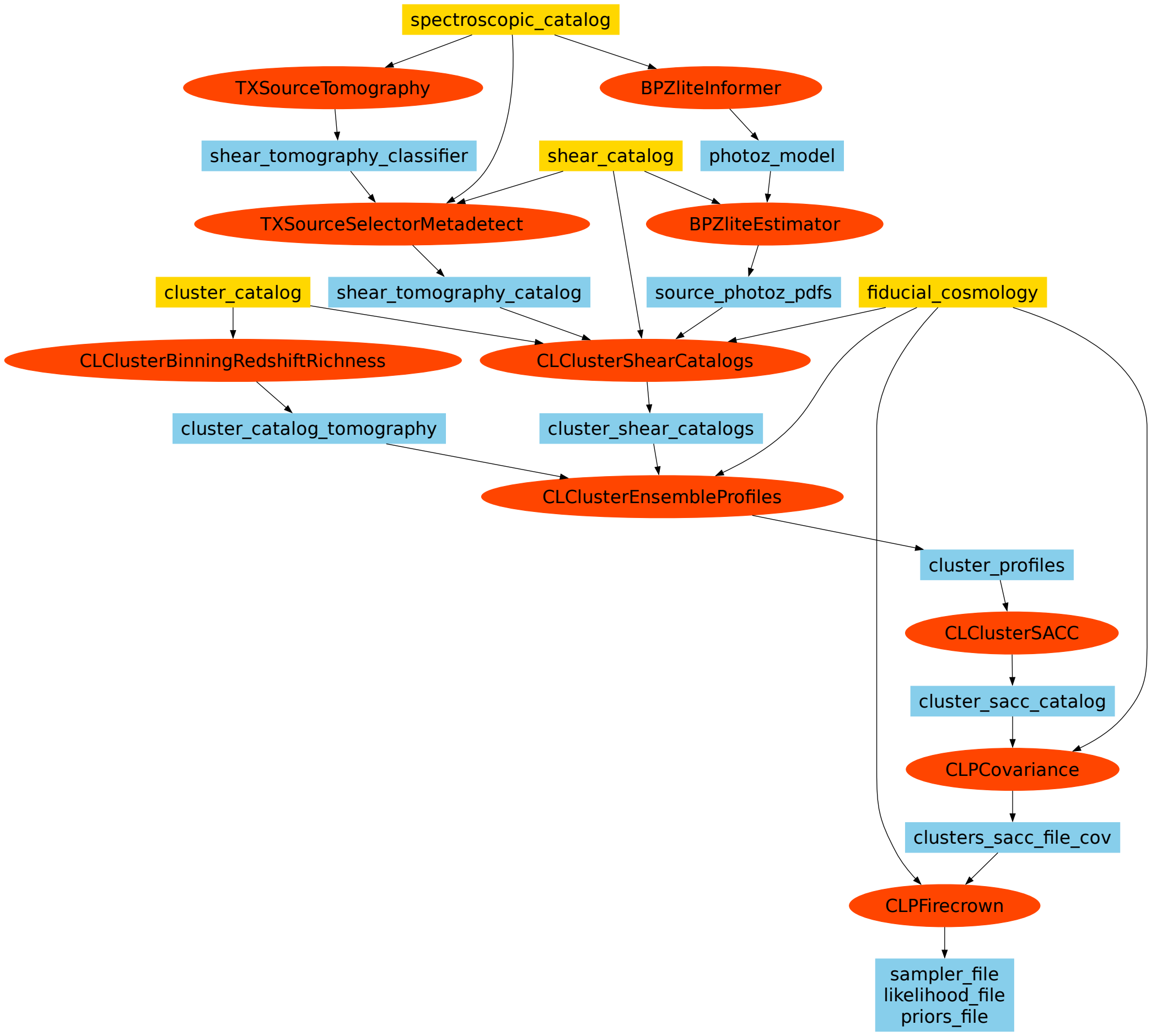}
    \caption{Flow chart of the \texttt{Ceci} pipeline structure described in
    Sec.~\ref{sec:ceci}, generated automatically with \texttt{Ceci}. Yellow rectangles denote the pipeline's overall
    input files; sky-blue rectangles denote its output files; red ellipses
    denote individual pipeline stages, labeled with their stage names (TX-- and
    BPZ-- prefixes for \texttt{TXPipe}/\texttt{RAIL} stages, CLCluster--
    for cluster-specific \texttt{TXPipe} stages, and CLP-- for \texttt{CLPipe} stages). Arrows indicate file-based dependencies between stages.}
        \label{fig:ceci_pipeline}
\end{figure*}


\section{The \desc Cluster Pipeline}
\label{sec:interface}
This section describes the structure of the \pipeline, from catalog ingestion to cosmological parameter inference. We first outline the main workflow and processing stages (Sec.~\ref{sec:workflow}), and then detail the pipeline management and configuration with \texttt{Ceci} (Sec.~\ref{sec:ceci}). Next, we describe its components (Secs.~\ref{sec:backends}--\ref{sec:firecrown}): \texttt{CCL} and \texttt{CLMM} backends, data vector construction with \texttt{TXPipe}, modeling with \texttt{CROW} and \texttt{TJPCov}, and inference framework with \texttt{Firecrown}. We also discuss the pipeline performance for our test case in Sec.~\ref{sec:performance}.
All the codes above and the pipeline development are done with the \texttt{Python} programming language.

\subsection{Workflow}
\label{sec:workflow}
The \pipeline aims to be an end-to-end framework for galaxy cluster cosmological analysis for LSST. The pipeline, hosted in the \texttt{CLPipe} repository, integrates a suite of DESC tools into a unified workflow managed by \texttt{Ceci}\footnote{\url{https://github.com/LSSTDESC/Ceci}}.

The pipeline, whose overall structure is illustrated in Fig.~\ref{fig:cluster_pipeline_diagram}, follows three main stages: (i) catalog ingestion and data vector estimation; (ii) covariance matrix evaluation; and (iii) likelihood construction and parameter inference. 

The workflow goes as follows: two configuration files define the stages to run and the analysis choices. The pipeline then begins with a catalog of detected galaxy clusters (giving at least their sky positions, the cluster redshifts, and the mass proxies) and a catalog of galaxies (giving at least their sky positions, shapes, redshift or photometry, and shear calibration information), representing the raw lensing dataset to assess the cluster lensing profiles.

These two catalogs are processed by \texttt{TXPipe}\footnote{\url{https://github.com/LSSTDESC/TXPipe}}~\citep{Txpipe}, which computes cluster abundance measurements and individual/stacked cluster weak lensing profiles in bins of mass proxy and redshift. Specifically for cluster lensing, \texttt{TXPipe} imports the cluster-related data management routines of \texttt{CLMM}\footnote{\url{https://github.com/lsstdesc/clmm}}~(\citealt{Aguena2021clmmv1, Aguena2026clmmv2}) to compute lensing profiles and their covariance matrices using data resampling techniques (jackknife, bootstrap, etc.). The output data vectors and data-driven lensing covariances are stored in the format handled by \texttt{SACC}\footnote{\url{https://github.com/LSSTDESC/sacc}}, the DESC standard for the storage of data vectors, covariances, and metadata. The \texttt{SACC} products are then read with \texttt{TJPCov}\footnote{\url{https://github.com/LSSTDESC/TJPCov}}, which computes theoretical covariance matrices for cluster abundance measurements using \texttt{CROW}\footnote{\url{https://github.com/LSSTDESC/crow}}. Finally, \texttt{Firecrown}\footnote{\url{https://github.com/LSSTDESC/firecrown}} performs likelihood evaluation and MCMC sampling, using the \texttt{SACC} files as input. It also combines the data vectors and covariances with theoretical predictions for cluster abundance and weak lensing profiles, computed using \texttt{CROW}, built upon the modeling capabilities of \texttt{CCL}\footnote{\url{https://github.com/LSSTDESC/CCL}}~\citep{2019ApJS..242....2C} and \texttt{CLMM}\footnote{\texttt{CLMM} can be either used to handle lensing data in cluster fields, or model individual and stacked cluster lensing profiles.}, to derive posterior constraints on cosmological and mass-observable parameters. Although the pipeline performs and saves the computation of unbinned cluster lensing profiles, the inference is currently developed only for a binned cluster analysis. 

The specific analysis configuration for LSST-DESC cluster cosmology is still to be defined, as it must be optimized against the upcoming LSST data and its systematics. The modular design of the pipeline's components makes it straightforward to extend or reconfigure the analysis as the science develops. Incorporating the unbinned likelihood, weak lensing mass, or a broader combination of cluster probes such as the 6$\times$2pt$+N$ analysis would not require redesigning the pipeline.

\subsection{Pipeline integration with \texttt{Ceci}}
\label{sec:ceci}

To run the pipeline, we use the \texttt{Ceci} pipeline manager package, which connects all of the analysis components discussed in the previous section. In Fig.~\ref{fig:ceci_pipeline}, we show a possible pipeline configuration for our baseline analysis presented on cosmoDC2 in Sec.~\ref{sec:results} as an example. Our implementation uses several \texttt{ceci.PipelineStage} objects\footnote{Each stage specifies its required inputs, outputs, and configuration parameters, with a \texttt{run()} method performing the corresponding computation. \texttt{Ceci} uses these declared inputs and outputs to connect the stages and determine their execution order.}, each responsible for a single operation in the analysis workflow, shown as red ellipses.

The stages in Fig.~\ref{fig:ceci_pipeline} come from three code bases. Stages with the \texttt{TX} prefix are general \texttt{TXPipe} stages, also used in other DESC analyses such as 3$\times$2pt. \texttt{CLCluster} stages are the cluster-specific measurement stages, developed for this work within \texttt{TXPipe}. Stages with the \texttt{BPZ} prefix belong to the \texttt{RAIL}\footnote{\url{https://github.com/LSSTDESC/RAIL}}~\citep{rail_pipeline} pipeline for redshift estimation. \texttt{CLP} stages were developed in \texttt{CLPipe} to connect the outputs of the previous stages to \texttt{TJPCov} and \texttt{Firecrown}. All stages are configured and run together by \texttt{CLPipe} as a single workflow. Stages are connected through input and output files, so execution order follows data dependencies rather than an explicit call sequence.


This file-based connection is what makes the pipeline modular. Any stage can be replaced with another that produces compatible outputs, and new stages can be added for extra computations, without changing the rest of the pipeline. The overall run is set through two YAML files: (i) one defining the \texttt{PipelineStage} objects to be used, their connections, and inputs/outputs paths; and (ii) a configuration file containing meta-parameters for each \texttt{PipelineStage}.

This second configuration file is where most of the modeling choices are set. As an example, for the \texttt{CLPFirecrown} module, the configuration file controls both the model choices (mass definition, halo mass function, scaling relation parameters) and the inference setup (which probe to use, priors, and sampler settings). Listing~\ref{lst:firecrown_config} shows a representative subset of this configuration\footnote{The full \texttt{CLPFirecrown} configuration includes several additional parameters, and a reduced subset is shown here for illustration.}. Changing a model choice, such as the halo mass function or mass definition, requires editing a single string in this file, with no changes to the pipeline code itself.

\begin{lstlisting}[float=tbp,language=yaml, caption={Example configuration of a \texttt{CLPipe} module.}, label={lst:firecrown_config}]
CLPFirecrown:
    hmf: 'despali16' 
    mass_def: '200c'
    survey_name: 'test_survey'
    use_shear_profile: true
    is_deltasigma: true
    use_cluster_counts: true
\end{lstlisting}

\texttt{Ceci} also enables MPI parallelization of computationally intensive stages, distributing them across multiple processes. The posterior sampling itself is performed subsequently using \texttt{CosmoSIS} with the \texttt{Firecrown} likelihood, executed outside the \texttt{Ceci} workflow, since the chains can run for hours or days and are often tested or restarted independently of the other stages.

\subsection{Backends: \texttt{CCL} and \texttt{CLMM}}
\label{sec:backends}
This section describes the two backends used in the pipeline. Both \texttt{CCL} and \texttt{CLMM} are existing DESC tools, not developed as part of this work.

The Core Cosmology Library (\texttt{CCL}) is the LSST DESC library that computes fundamental cosmological quantities, supporting many cosmologies in addition to $\Lambda$CDM. In this pipeline, \texttt{CCL} is not used as a standalone analysis stage, but rather as a common cosmology engine accessed consistently by all major modules, including \texttt{TXPipe}, \texttt{CLMM}, \texttt{CROW}, \texttt{TJPCov}, and \texttt{Firecrown}, ensuring internal consistency across all stages of the pipeline.

In this work, we restrict the analysis to $\Lambda$CDM. Extensions whose parameters are already supported by \texttt{CCL} and \texttt{Firecrown}, such as $w_0w_a$CDM, can be sampled by adding them to the \texttt{CLPFirecrown} configuration, although their propagation through all modules remains to be validated. Since \texttt{CCL} already supports several other cosmological models and serves as the shared cosmology engine for every major module, extending the pipeline to other models would mainly involve configuring or extending \texttt{CCL} itself. The theory modules (\texttt{CROW}, \texttt{TJPCov}, \texttt{CLMM}), which already rely on \texttt{CCL} for their cosmological calculations, would then need to be checked for consistency, while \texttt{Firecrown} would additionally need to support sampling the parameters of the new cosmological model.

The Cluster Lensing Mass Modeling (\texttt{CLMM}) library is the weak lensing backend of the pipeline, used both at the data level within \texttt{TXPipe} (Sec.~\ref{sec:txpipe}) and at the theory level within \texttt{CROW} (Sec.~\ref{sec:crow}), and it relies on \texttt{CCL} for cosmological calculations\footnote{\texttt{CCL} is one of \texttt{CLMM}'s cosmology backends, along with \texttt{NumCosmo}. \texttt{CCL} is used by default.}. Here we use several features introduced in version 2 of \texttt{CLMM}, which was developed to include new implementations supporting the first weak-lensing analyses of LSST clusters. Details of \texttt{CLMM} v2 are presented in \citet{Aguena2026clmmv2}.

\subsection{Data vector construction with \texttt{TXPipe}}
\label{sec:txpipe}

In the \pipeline, \texttt{TXPipe} performs the measurements needed for the cluster abundance and weak lensing analyses. Below we describe the \texttt{TXPipe} stages grouped in different blocks. The first block uses existing stages from the general \texttt{TXPipe} framework (\texttt{TX}-- stages, together with the \texttt{RAIL} \texttt{BPZ}-- stages in Fig.~\ref{fig:ceci_pipeline}), while the cluster abundance and weak lensing measurement stages, along with the output SACC format, were developed as part of this work (\texttt{CLCluster}-- stages). \\

\noindent\textit{Data ingestion, galaxy–redshift assignment, and shear calibration:} This first block follows the framework described in \citet{Txpipe}. It begins by ingesting pre-processed galaxy and cluster catalogs that will be saved in the HDF5 data format, which may include multi-band photometry, shape measurements, and associated calibration information. The cluster catalog contains positions, observable properties (e.g., richness, a count of member galaxies typically weighted by their membership probabilities), identifiers assigned by the cluster finder, and redshifts. Following ingestion, \texttt{TXPipe} assigns redshifts to galaxies either by estimating them internally from photometry using \texttt{RAIL} or from external inputs. When redshift point estimates are provided, \texttt{TXPipe} uses these directly. Otherwise, the pipeline operates on full photometric redshift probability distributions. Finally, the galaxy catalog undergoes selection cuts in redshift and signal-to-noise ratio, with shear response calibration and selection response corrections applied following the Metadetection and Metacalibration formalism~\citep{metacal}, depending on catalog type. \\

\noindent\textit{Cluster abundance and weak lensing measurements:}
For cluster weak lensing, the measured quantity is the ellipticity of each source galaxy, whose tangential component around a cluster traces the shear induced by the cluster's projected mass~\citep{Bartelmann2001}. Averaging this tangential component in radial bins around the cluster center gives the cluster weak lensing profile. Building this profile requires source galaxies' redshift information, used both to select galaxies behind the cluster and to set the lensing efficiency of each cluster--source pair~\citep{mandelbaum18}.

When individual source redshifts are available, we identify source galaxies around each cluster using their individual redshift information and a neighbor search within a maximum angular radius. This maximum is set per cluster by converting a fixed physical search radius into an angle at the cluster's redshift, using a fiducial cosmology specified in the configuration file and computed with \texttt{CCL}. The sensitivity of the measured data vectors to this cosmological choice is explored in App.~\ref{app:ds_profiles_diff_fiducial}. We are also implementing a tomographic bin-based source selection for the case where only redshift bin membership is known, avoiding the need for individual photometric redshift point estimates in future analysis.

We compute the cluster weak lensing probes using \texttt{CLMM} as the underlying library. The pipeline currently supports two lensing probes: the reduced tangential shear ($g_t$), the tangential distortion of background galaxy shapes induced by the cluster's gravitational potential, and the excess surface density ($\Delta\Sigma$), the shear rescaled by the critical surface density of each cluster--source pair. We refer the reader to \citet{Aguena2026clmmv2} for the full estimator formalism.

We compute individual cluster profiles from the cluster-galaxy pairs, providing unbinned lensing measurements for each cluster. In parallel, we bin clusters in redshift and cluster observable, which directly gives the cluster abundance in each bin. For this same binning scheme, we radially bin the individual profiles and combine them to construct the stacked weak lensing profiles. We then apply \texttt{CLMM}'s bootstrap or jackknife resampling methods to derive a data-driven covariance of the stacked lensing profiles.\\

\noindent\textit{Output products:}
At the end of the \texttt{TXPipe} framework, in \texttt{CLClusterSACC}, we write both cluster counts and stacked weak lensing profiles per bin to a \texttt{SACC} file, together with their covariance matrices and binning information. For cluster counts, the covariance is just the number of counts $N_{ij}$ in each bin, saved as a placeholder in \texttt{SACC}, to be replaced by the full theoretical covariance in the next step of the pipeline. We also save individual cluster lensing profiles as pipeline outputs but these are not currently included in the \texttt{SACC} file.

\begin{figure*}[tbp]
    \centering
    \includegraphics[width=1\linewidth]{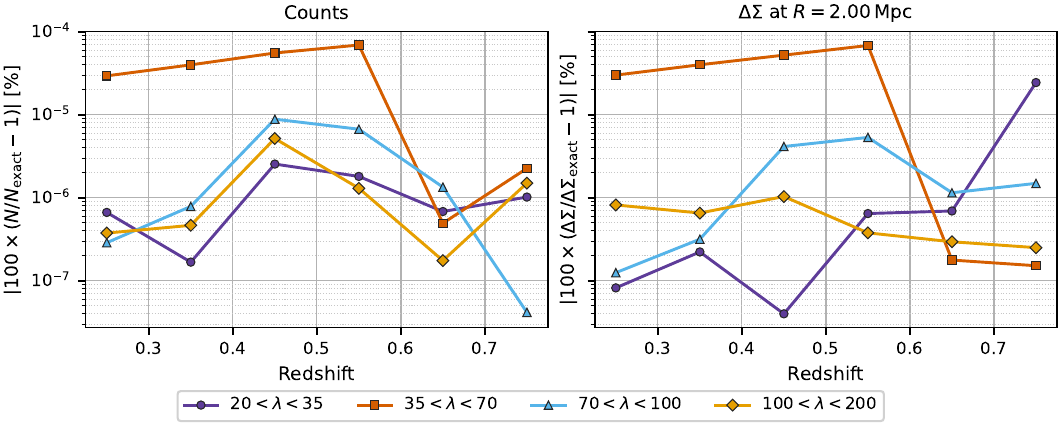}
    \caption{Left: precision plot for the \texttt{CROW} count prediction comparing tabulated to non-tabulated results, assuming the latter to be the truth. Right: Same as left, but for $\Delta\Sigma(R=2\ {\rm Mpc})$. Colors and markers denote the richness bins.}
    \label{fig:precision_plot}
\end{figure*}

\subsection{\texttt{CROW}: Theoretical prediction for cluster number counts and stacked lensing profiles}
\label{sec:crow}

We developed the Cluster Reconstruction of Observables Workbench (\texttt{CROW}), a DESC tool for computing theoretical predictions of galaxy cluster probes. In practice, it combines the halo mass function, mass-observable relation, and selection function within a halo-model framework to predict cluster counts and cluster weak lensing profiles using \texttt{CCL} and \texttt{CLMM}.  It is used both in \texttt{TJPCov} for cluster counts covariance calculation and in \texttt{Firecrown}, where its predictions are compared against the data vectors at the likelihood level.

Formally, for bins of redshift and cluster observable $(i,j)$, this halo-model framework predicts the number counts and stacked weak lensing profile as
\begin{align}
\label{cluster_abundance}
    N_{ij} &= \Omega_S\int\limits_{z_i}^{z_{i+1}} dz \int\limits_{\ln x_j}^{\ln x_{j+1}}d\ln x  \int\limits_{M_{\rm min}}^{M_{\rm max}} dM \times \nonumber \\
    &\frac{d^2V(z)}{dzd\Omega}\, \frac{dn(M,z)}{dM}\,P\left(\ln x\mid M, z\right) \,\Phi(M,x,z)
,\end{align}
\begin{align}
\label{cluster_theta_profile}
    &\Theta_{ij}(R) = \frac{\Omega_S}{N_{ij}}\int\limits_{z_i}^{z_{i+1}} dz \int\limits_{\ln x_j}^{\ln x_{j+1}}d\ln x  \int\limits_{M_{\rm min}}^{M_{\rm max}} dM \times \nonumber \\
    &\frac{d^2V(z)}{dzd\Omega}\, \frac{dn(M,z)}{dM}\,P(\ln x\mid M, z)\, \Phi(M,x,z)\,\Theta(R|M)
,\end{align}
where the integrals run over (i) $z$, the cluster's redshift, (ii) $M$, the halo mass\footnote{Since there are no halos above and below some mass thresholds, the mass integral is performed within a minimum and maximum instead of $\pm \infty$ for computational purposes.}, and (iii) $x$, the cluster observable, related to the halo mass through the mass-observable relation $P(\ln x\mid M,z)$. We made no distinction between the halo and associated cluster redshifts, though this could be accounted for by marginalizing over the cluster's redshift estimate and will be done in the future. $\Omega_S$ represents the survey footprint area in steradians~\citep{Lacasa2018}, $d^2V(z)/dzd\Omega$ is the comoving volume element, $dn(M,z)/dM$ is the halo mass function, denoting the mass and redshift distribution of the dark matter halos (e.g.~\citealt{Tinker2008, despali2016universality}), whose most massive ones host clusters of galaxies, and $\Phi(M,x,z)$ is the selection function, accounting for the cluster selection systematics. In Eq.~\eqref{cluster_theta_profile}, $\Theta$ denotes the weak lensing probe, ($\Delta\Sigma$, $g_t$), consistent with the two probes supported by our pipeline (Sec.~\ref{sec:txpipe}). The code also supports computation of the mean halo mass in each bin, $\langle M\rangle_{ij}$, obtained by replacing $\Theta(R|M)$ with $M$ in Eq.~\eqref{cluster_theta_profile}.

Each ingredient entering Eqs.~\eqref{cluster_abundance} and~\eqref{cluster_theta_profile} is instantiated in \texttt{CROW} as a separate, modular object: a \texttt{CROW.ClusterAbundance} or \texttt{CROW.ClusterShearProfile} theory object (both built from a \texttt{CCL} cosmology and halo mass function), a mass-observable relation object implementing $P(\ln x\mid M,z)$, selection function objects jointly forming the selection function $\Phi(M,x,z)$, and a redshift object relating the observed cluster redshift to the true halo redshift, either directly (no integral) or through a photometric redshift kernel (future work). \texttt{CROW.ClusterShearProfile} additionally controls every lensing-specific choice, the predicted probe ($\Delta\Sigma$ or $g_t$), halo concentration, one- vs. two-halo contributions, boost factor corrections, and miscentering. For reduced shear specifically, it also sets the lensing-efficiency approximation and whether it is interpolated. These objects are combined by a \texttt{CROW.recipe} object, which performs the integrals described above.

Currently, the halo mass functions are computed via \texttt{CCL}, the mass-observable relation implements a logarithmic mass-richness relation as in \cite{Murata2019}, Eq.~\eqref{mass_richness_relation}, and the selection function is modeled through completeness (missed detections) and purity (false detections) terms as in~\cite{Aguena2018}, Eqs.~\eqref{selection1}--\eqref{selection2}. Since each ingredient above is a standalone object, any of these can be altered independently\footnote{For instance, we can employ other mass-observable relations such as SZ signal-to-noise~\citep{ade2016planck} or X-ray luminosity~\citep{vikhlinin2009chandra}, or alternative selection function parametrizations, without changes to the overall framework.}, and other parametrizations will be tested in future analyses.

\texttt{CROW} supports two modes for evaluating the multi-dimensional integrals in Eqs.~\eqref{cluster_abundance} and~\eqref{cluster_theta_profile}: a fast, tabulated mode using precomputed grids and Simpson's rule from \texttt{SciPy}~\citep{2020SciPy-NMeth} (\texttt{CROW.GridBinnedClusterRecipe}), and a direct, non-tabulated mode using the \texttt{NumCosmo} integration framework~\citep{dias2014numcosmo} (\texttt{CROW.ExactBinnedClusterRecipe}). The tabulated mode is $\sim$100$\times$ faster while maintaining accuracy. We show in Fig.~\ref{fig:precision_plot} the relative difference between the \texttt{CROW} tabulated approach and non-tabulated results, reaching $\sim10^{-5}\%$ for both counts and lensing profiles. For this benchmark in Fig.~\ref{fig:precision_plot}, we used binned abundance plus stacked $\Delta\Sigma$ predictions with: \cite{Murata2019} scaling relation, \cite{despali2016universality} halo mass function; \cite{Aguena2018} selection function; 4 redshift bins $z \in[0.2, 0.8]$ and 5 richness bins $\lambda \in [20, 200]$; 10 radial bins $R \in [1.0, 10]$\,Mpc for lensing profiles; grid resolution of 40 points in mass with $M \in [10^{12}, 10^{15.5}] M_\odot$, and 20 points for each redshift and richness bins respectively. The tabulated mode is recommended for inference while the non-tabulated mode remains available for validation.

\subsection{Theoretical covariance with \texttt{TJPCov}}
\label{sec:tjpcov}
\texttt{TJPCov} is DESC's general-purpose covariance calculator. We developed the theoretical cluster count covariance within \texttt{TJPCov}, using \texttt{CROW} and \texttt{CCL} as backends, which we describe below. This covariance matrix $\mathrm{Cov}_{N}$ is given by
\begin{align}
\label{ssc_cov}
    [\mathrm{Cov}_{N}]_{ijkl} &= \delta^K_{ik}\delta^K_{jl}N_{ij} + [\mathrm{Cov}_{\rm SSC}]_{ijkl}  
,\end{align}
where $i,k$ run over redshift bins and $j,l$ over cluster observable bins ($\delta^K_{ik}$ is the Kronecker symbol). In Eq.~\eqref{ssc_cov}, the first term is the Poisson shot-noise variance contribution, since it is a count experiment. The second term is the Super Sample Covariance (SSC), arising from fluctuations of modes larger than the survey volume~\citep{Hu2003,Lacasa2018, fumagalli2021euclid}, and which couples the local halo number densities across space and mass–redshift bins. The SSC term adds variance and correlation between the abundance of galaxy clusters measured in different redshift-richness bins and is given by~\citep{Lacasa2018}
\begin{align}
\label{ssc_contribution}
    [\mathrm{Cov}_{\rm SSC}]_{ijkl} &= N_{ij} N_{kl} \langle b\rangle_{ij} \langle b\rangle_{kl} S_{ik}
,\end{align}
where $\langle b\rangle_{ij}$ is the halo bias averaged over the cluster population in the $(i,j)$ redshift--observable bin, and $S_{ik}=\langle\delta_{m,i}\delta_{m,k}\rangle$ is the covariance of the survey-window-smoothed matter overdensity between the $i$-th and $k$-th redshift bins. Compared to the Poisson shot noise, the SSC contribution is proportional to $N_{ij}^2$, which makes it non-negligible for LSST-sized cluster catalogs \citep{Payerne2023,Payerne2024}. 

We use \texttt{TJPCov} within \texttt{CLPipe} in the \texttt{CLPCovariance} stage (Fig.~\ref{fig:ceci_pipeline}), where the \texttt{SACC} file produced by \texttt{TXPipe} is used to extract the binning and survey area information of the data vectors.  In addition, fiducial cosmological parameters and fiducial mass-observable relation parameters are provided as inputs to compute the theoretical cluster counts covariance.

The resulting theoretical covariance matrix replaces the data-driven cluster-count covariance in the \texttt{SACC} file, while the weak-lensing covariance provided by \texttt{CLClusterEnsembleProfiles} is left unchanged. A theoretical covariance model for cluster weak lensing is currently under development.

\subsection{Likelihood-based inference with \texttt{Firecrown}}
\label{sec:firecrown}

\texttt{Firecrown} is DESC's likelihood framework, providing the infrastructure for implementing likelihoods. In the \pipeline, it is the final stage combining data and theoretical predictions for Bayesian parameter inference. Within the existing \texttt{Firecrown} structure, we developed statistical objects and their likelihood files for the binned cluster abundance and stacked weak lensing profile (an unbinned, individual-profile likelihood is left for future work). All cluster-related \texttt{Firecrown} configuration files are generated at the \texttt{CLPFirecrown} stage of \texttt{CLPipe}. We describe the resulting \texttt{Firecrown} flow below.\\

\noindent\textit{Data and parameter input:}
In \texttt{Firecrown}, the binned cluster counts and stacked weak lensing statistic objects read the \texttt{SACC} file produced by \texttt{TJPCov}, loading the data vector and associated metadata. The covariance itself is read using \texttt{Firecrown}'s existing infrastructure. The parameters to be constrained, including cosmological parameters, mass-observable relation parameters, and others, are specified along with their prior distributions in configuration files.\\

\noindent\textit{Likelihood:} 
We create a \texttt{Firecrown} likelihood \texttt{Python} file that combines both probes using our developed statistics. Cluster abundance and the cluster lensing profile measurement are correlated, as they arise from the same underlying matter density field and share some systematics. It is generally assumed, however, that any cross-covariance between them is negligible compared to the proper covariance terms, as justified in~\cite{Costanzi2019}. We adopt this assumption in the current implementation, which can be extended to a joint likelihood including the cross-correlation between the two probes. The likelihood then reads
\begin{align}
\label{likelihood_comb}
\mathcal{L}_{\rm joint}(\{\widehat{N}_{ij}\},\{\widehat{\Theta}_{ij}\}) 
=
\mathcal{L}_{N}(\{\widehat{N}_{ij}\})
\times
\mathcal{L}_{\Theta}(\{\widehat{\Theta}_{ij}\}).\end{align}
In the above, the hat overscript relates to the measured quantities, while its absence denotes their predictions. The abundance likelihood $\mathcal{L}_N$ is described with a Poisson likelihood for small samples, and with a Gaussian likelihood once the super-sample covariance becomes comparable to the shot noise~\citep{Payerne2024}. In our binned implementation, we adopt the latter, modeled as a multivariate Gaussian across all redshift-richness bins~\citep{Hu2003, Payerne2023}, using the covariance matrix of Eq.~\eqref{ssc_cov}. We similarly adopt a Gaussian likelihood for the cluster weak lensing probe $\Theta \in \left\{\Delta\Sigma, g_t\right\}$~\citep{McClintock2019mass, desclustercosmo20}. At each point in parameter space, \texttt{Firecrown} updates all sampled parameters, with cosmological parameters specifically propagated through its interface with \texttt{CCL}. Our statistical objects then evaluate the theoretical predictions for cluster abundance and weak lensing profiles using \texttt{CROW}. For the moment, the covariance is held fixed throughout inference.

\texttt{Firecrown} further interfaces with several statistical frameworks, including \texttt{Cobaya}~\citep{torrado2021cobaya}, \texttt{NumCosmo}, and \texttt{CosmoSIS}~\citep{zuntz2015cosmosis}, each providing its own set of native samplers and cosmological backends, allowing the use of different sampling algorithms and parameterizations. The pipeline integration framework generates all files required for the inference, while the final execution of the sampler is performed separately.\\

\noindent\textit{Bayesian inference and output:}
Once all configuration files are in place, we run the MCMC inference with the \texttt{Firecrown} likelihood and the chosen sampler, producing posterior chains from which parameter constraints and posterior covariances are derived.

\subsection{Computing performance}
\label{sec:performance}

The \pipeline has been tested at the National Energy Research Scientific Computing Center (NERSC), DESC's primary computing facility, and at the IN2P3 Computing Center (CC-IN2P3), which hosts the Rubin Observatory French Data Facility. We discuss its performance on the current implementation validated on simulations covering $440\, \mathrm{deg}^2$ of the sky (see Sec.~\ref{sec:results}). The long-term goal is a pipeline that scales to Y10 LSST data volumes, expected to be $40$ times bigger than this test case.

\needspace{3\baselineskip} 
\subsubsection{Data processing}
Within \texttt{TXPipe}, we load large catalog files in chunks rather than reading entire files into memory, a task made straightforward by the HDF5 format. MPI parallelization, managed by \texttt{Ceci}, distributes computations per chunk to reduce wall-clock time. In the specific case of our analysis, most output files (\texttt{Metadetect} shear catalog, cluster catalog, lensing profiles, source photo-$z$ probability distributions) have modest storage footprints, ranging from $2$ to $80$ GB for the 36 million galaxies selected. However, scaling such data products to Y10 LSST statistics remains an open challenge. \texttt{TXPipe} requires a peak of $\sim$40~GB of RAM in parallelized mode and completes in approximately 10 minutes for our test case\footnote{Note that the data processing will not be performed several times once there is a stable version.} ($\sim$16 minutes when photo-$z$ estimation with \texttt{RAIL} is included, and $\sim$1.5 hours with source catalog ingestion). This computation time depends strongly on parallelization and on the number of galaxies carried through the pipeline at each stage, set first by the ingest step and again by the source selection both in the shear calibration stage and at the cluster source pairing level. Assessing the computational requirements of repeated catalog production and validation at LSST-DESC scale remains an area for future work.

One current limitation of the \texttt{Ceci} pipeline manager is uniform MPI resource allocation: requesting 60 MPI jobs per CPU applies that allocation to all stages, even those requiring no parallelization, leaving resources underutilized. The overhead is negligible, and the results are unaffected, but this represents suboptimal resource use. Addressing this through dynamic job allocation via \texttt{Parsl} is under study.

\subsubsection{Computing theoretical prediction and inference} 

For the given analysis binning scheme, the count covariance calculation with \texttt{TJPCov} runs in $\sim$1 minute with no memory bottlenecks. However, recomputing the covariance at every MCMC iteration is currently impractical. Developing a more dynamic covariance evaluation scheme is an important direction for future development. The covariance is computed once at the desired fiducial cosmology and held fixed.

Then, the primary computational bottleneck lies in the theoretical prediction stage. In the baseline analysis, the prediction of galaxy cluster probes (counts and $\Delta\Sigma$) involves nested three-dimensional integrals (Eqs.~\eqref{cluster_abundance} and~\eqref{cluster_theta_profile}). When evaluated directly using numerical integration routines within \texttt{CROW}, it requires in total $\sim$20~seconds, making MCMC analyses computationally impractical. This evaluation time is expected to grow when adding more integrals (e.g., by marginalizing over cluster photometric redshift uncertainties).

To tackle this problem, we adopt a grid-based implementation of the nested integrals. For the baseline analysis presented in Sec.~\ref{sec:results}, the total computation of the \texttt{Firecrown} likelihood evaluation requires $\sim 0.7\,\mathrm{s}$, with the \texttt{CAMB} power spectra computation contributing a further $\sim 1.5$ s\footnote{Here we use \texttt{CAMB} as the initial computation, passing background and linear power spectrum quantities to \texttt{pyccl}. We disable the CMB Boltzmann solve to speed up the code, verifying these choices leave the likelihood numerically unchanged.}. Parallelizing across MPI workers matched to the sampler's ensemble batch size, $\sim$300,000 posterior samples are generated in $\sim$7 hours. 

Moreover, at the inference stage, parallelization is available for \texttt{CosmoSIS}-based runs, though \texttt{Firecrown} computations are currently serial. Parallelizing the likelihood evaluation itself offers a natural avenue for future optimization if the parameter space expands.

\begin{figure*}[tbp]
    \centering
    \includegraphics[width=\textwidth]{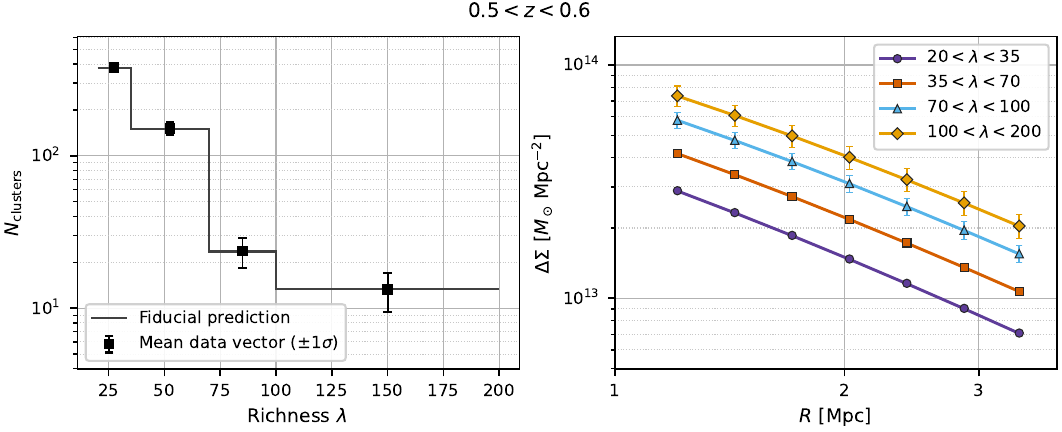}
    \caption{Validation of the mock data vectors against the pipeline's
    theoretical predictions, for the representative redshift bin
    $0.5 < z < 0.6$ and all richness bins. \textit{Left}: cluster counts
    per richness bin, comparing the mean over the $1\,000$ mock
    realizations (points) with the fiducial \texttt{CROW} prediction
    evaluated at the input parameters of Table~\ref{tab:fiducial} (step lines). \textit{Right}: same comparison for the stacked excess
    surface density profiles $\Delta\Sigma(R)$, with the mean mock data vectors as points and the fiducial predictions as lines, one per richness bin. Error bars show the $1\sigma$ uncertainty of a single realization, from the covariance of the $1\,000$ mocks.
    The mean mock data vectors agree with the fiducial predictions at
    the percent level. The other redshift bins show a similar level of
    agreement.}
    \label{fig:capish_mean_agreement}
\end{figure*}

\section{Validation with synthetic mock datasets}
\label{sec:mock_analysis}
Before applying the \pipeline to realistic simulations, we first test its inference machinery on mock data for which the underlying model is exactly known. Generating mocks with the same model we will use for the likelihood inference isolates any bias arising from the inference chain itself (\texttt{CROW}, \texttt{TJPCov}, and \texttt{Firecrown}) from systematics associated with real data. If the pipeline cannot recover the input parameters when the underlying model is known exactly, no result built on top of it can be trusted. 

We perform this test using the same cluster counts and stacked excess surface density ($\Delta\Sigma$) analysis, binned in redshift and richness, presented in Sec.~\ref{sec:results}, and show that the fiducial cosmological and mass-observable parameters are recovered consistently when the analysis model matches the data-generation model. Note that this test does not exercise the data measurement part of the pipeline, mainly \texttt{TXPipe}, since mock data vectors are obtained directly from the mock generation code\footnote{Passing mock catalogs through the full \texttt{TXPipe} measurement chain would require assigning realistic sky positions, generating full multi-band photometry catalogs, and inserting placeholder calibration columns, unneeded here and unnecessarily complicated.}. We describe the mock data generation and its agreement with the pipeline's theoretical predictions in Sec.~\ref{mock_generation} and present the resulting parameter constraints in Sec.~\ref{cosmo_constraints_mocks}.

\subsection{Synthetic mock data generation}
\label{mock_generation}

To produce the binned mock data vectors, we need realistic cluster and galaxy source catalogs. Directly predicting the binned counts and lensing profiles with \texttt{CROW} would not be sufficient, as it bypasses the unbinned, per-halo and per-galaxy scatter that gives rise to a realistic data-vector covariance in the first place. We therefore generate mock catalogs at the unbinned, individual level, using the Python code \texttt{capish}~\citep{payerne2026simulation} together with \texttt{CLMM}, before binning the resulting profiles and counts identically to the baseline analysis (see Sec.~\ref{sec:results}).

Using \texttt{capish}, we generate the underlying halo population (true masses $M_{\rm 200c}$ and redshifts $z_{\rm true}$) over 436.6 deg$^2$ through an SSC log-normal and Poisson sampling of the \citet{despali2016universality} halo mass function at the unbinned level. Cluster richnesses $\lambda_{\rm obs}$ are then generated as log-Gaussian sampling of the mass-richness relation~\citep{Lima2005, Murata2019}
\begin{align}    \label{mass_richness_relation}
    &P(\ln\lambda\mid M, z)  \propto \frac{1}{\sigma_{\ln \lambda}}\exp\left\{-\frac{1}{2}\left(\frac{\ln \lambda -\mu_{\ln \lambda}}{ \sigma_{\ln \lambda}}\right)^2\right\}
,\end{align}
where the mean cluster's richness $\lambda$ is linked to $M$ via a power law relation. We define the mean and standard deviation richness $\mu_{\ln \lambda}$ and $\sigma_{\ln \lambda}$, such that
\begin{align}
        \label{mean_scaling}
    & \mu_{\ln \lambda} = \mu_{0} +  \mu_{m} \ln\left(\frac{M}{M_p}\right) + \mu_{z}\ln\left(\frac{1+z}{1+z_p}\right),  \\
    \label{sigma_scaling}
    &\sigma_{\ln \lambda} =   \sigma_{0} + \sigma_{m} \ln \left( \frac{M}{M_p}\right) + \sigma_{z}\ln\left(\frac{1+z}{1+z_p}\right).
\end{align}
In the above, $M_p$ and $z_p$ are pivot values, equal to the ones in Table~\ref{tab:fiducial} caption. 
The mean relation is parametrized by an offset richness $\mu_{0}$, a mass dependence $\mu_{m}$, and a redshift dependency $\mu_{z}$, while the scatter is modeled through the analogous parameters $\sigma_{0}, \sigma_{m}, \sigma_{z}$. For simplicity, we do not apply any cluster finder selection effects (purity and/or completeness) on the resulting cluster catalog. 

For the galaxy sources, we use the mock generation modules of \texttt{CLMM} to (i) generate background galaxy catalogs\footnote{Using $\sigma_\mathrm{SN} = 0.0$, $z_{\rm gal} > z_{cl} + 0.2$, $\text{n}_{\rm gal} = 25\ \text{arcmin}^{-2}$, following \cite{Chang2013} galaxy redshift distribution.} for each galaxy cluster; (ii) lens these galaxies with a spherical NFW~\citep{nfw} profile evaluated at the true mass $M_{\rm 200c}$, redshifts $z_{\rm true}$ and concentration $c_{\rm 200c}=3.8$; (iii) estimate the perfectly-centered individual cluster excess surface density profile in physical radial bins; and (iv) compute stacked lensing profiles in the same richness-redshift bins as the baseline \redmapper analysis. Mock galaxy cluster abundance is then measured in the same bins. 

Since the mock data generation is stochastic, the level of agreement between a single realization's data vector and the theoretical prediction depends strongly on the specific random seed used. We therefore generate $1\,000$ independent mock realizations and compute the mean cluster counts and stacked lensing profiles across them. Fig.~\ref{fig:capish_mean_agreement} shows the percent-level agreement between the mean mock data vector and the fiducial prediction using \texttt{CROW}, for one representative redshift bin across all richness bins. Next, we will use this mean data vector to perform the parameter inference.

\begin{figure*}[p]
\centering\includegraphics[width=1.0\textwidth]{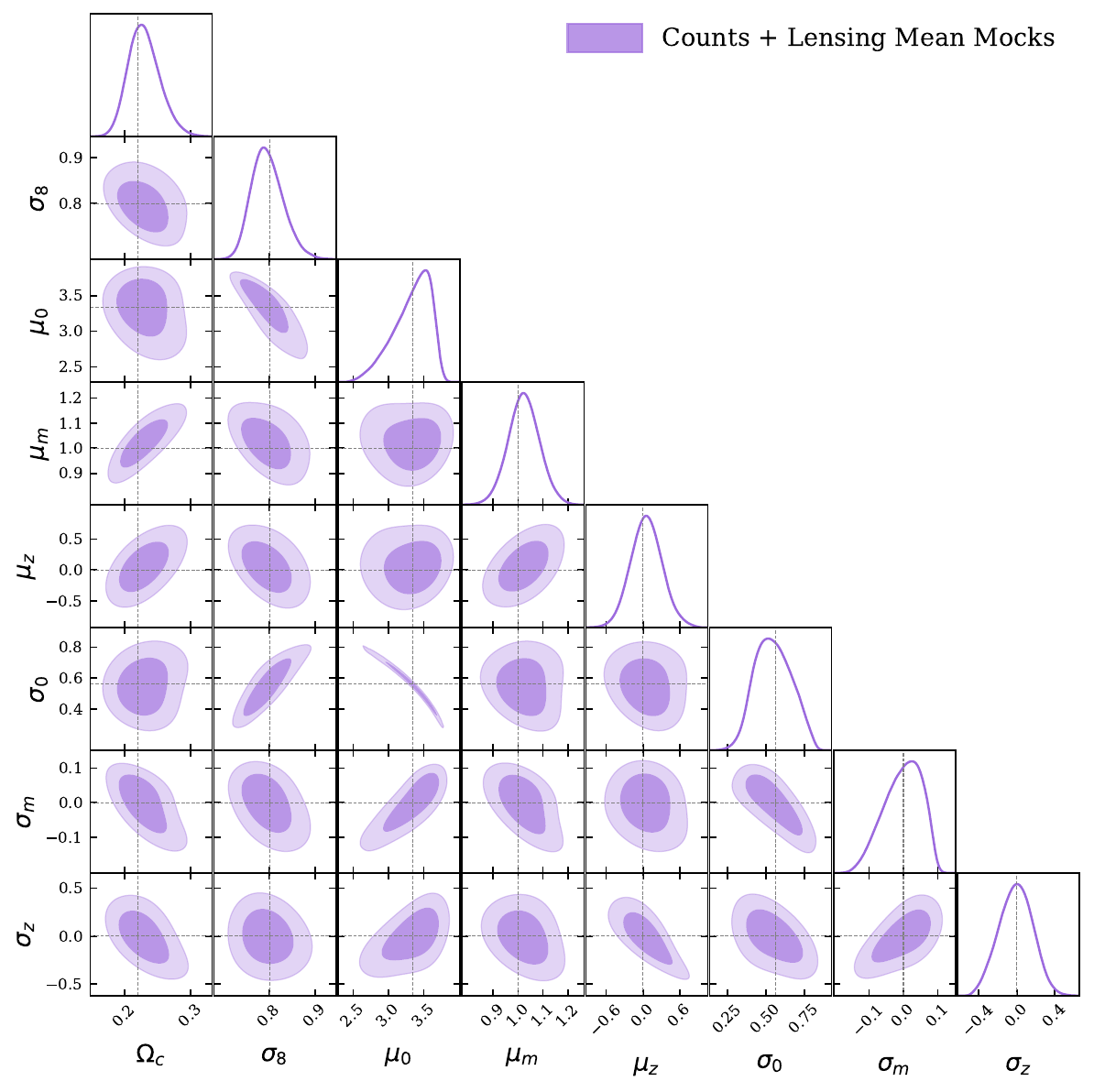}
    \caption{Posterior constraints on the six mass--richness relation parameters and the cosmological parameters $\Omega_c$ and $\sigma_8$, for the mean mock data vector. Contours show the 68\% and 95\% credible regions. Dashed lines indicate the fiducial cosmology and mass--richness parameters used to generate the mock catalog.}
\label{fig:capish_redmapper}
\end{figure*}

\subsection{Parameter posterior inference}
\label{cosmo_constraints_mocks}

We perform a joint analysis of cluster abundance and stacked cluster lensing profiles with a combined Gaussian likelihood as described in Sec.~\ref{sec:firecrown}. The counts covariance is assumed to be the theoretical Poisson+SSC covariance of Eq.~\eqref{ssc_cov} evaluated at the fiducial parameters. The lensing covariance entering this likelihood is estimated directly from the scatter of the $1\,000$ individual mock realizations about their mean, without further rescaling by the number of realizations. Pairing this single-realization-sized covariance with the (noise-free) mean data
vector follows the Asimov-dataset approach of
\citet{cowan2011asymptotic}. It tests whether the fiducial parameters are recovered with a single-realization-sized posterior.

We sample the joint posterior distribution of an eight-dimensional parameter space ($\mu_0, \mu_m, \mu_z, \sigma_0, \sigma_m, \sigma_z$, $\Omega_c$, $\sigma_8$) with the \texttt{emcee} sampler within the \texttt{CosmoSIS} framework\footnote{This mimics a real analysis where the scaling relation parameters are unknown and degenerate with cosmology.}, with over $2.8\times 10^6$ samples and the first  15\% discarded as burn-in. The priors and fiducial values adopted for the sampled parameters are shown in Table~\ref{tab:fiducial}.

\begin{table}[tbp]
    \centering
    \caption{Priors and fiducial values for the sampled parameters, for
    pivot values $M_\mathrm{pivot} = 10^{14.3}\,M_\odot$ and
    $z_\mathrm{pivot} = 0.5$. Second column: uniform prior distributions,
    shared between the idealized-mock analysis (Sec.~\ref{sec:mock_analysis}) and the cosmoDC2 analysis (Sec.~\ref{sec:results}), in addition to $\sigma_{\ln \lambda} > 0$.
    Third column: injected values used to
    build the idealized mocks. Fourth column: fiducials obtained from posterior means from the
    MCMC fit to the matched \redmapper cosmoDC2 halo catalog (see App.~\ref{appA}) and from the cosmoDC2 fiducial cosmology.}
    \begin{tabular}{lccc}
        \hline\hline
        Parameter & Prior & Mock & CosmoDC2\\
        \hline
        $\Omega_c$  & $\mathcal{U}(0.10, 0.50)$ & $0.22$  & $0.22$    \\
        $\sigma_8$  & $\mathcal{U}(0.5, 1.1)$   & $0.80$  & $0.80$    \\
        $\mu_0$     & $\mathcal{U}(2, 10)$      & $3.34$  & $3.3446 \pm 0.0131$    \\
        $\mu_m$     & $\mathcal{U}(0.5, 2.0)$   & $1.0$ & $0.9587 \pm 0.0225$       \\
        $\mu_z$     & $\mathcal{U}(-2, 2)$      & $0.0$     & $-0.0449 \pm 0.1160$   \\
        $\sigma_0$  & $\mathcal{U}(0.1, 2.0)$   & $0.56$  & $0.5626\pm 0.0098$         \\
        $\sigma_m$  & $\mathcal{U}(-0.6, 0.3)$  & $0.0$     & $0.0454 \pm 0.0092$       \\
        $\sigma_z$  & $\mathcal{U}(-0.5, 2.0)$  & $0.0$     & $-0.0458 \pm 0.0614$         \\
        \hline
    \end{tabular}
    \label{tab:fiducial}
\end{table}

We show in Fig.~\ref{fig:capish_redmapper} the posterior distribution of cosmology and scaling relation parameters, along with the fiducial values. Using the mean data vector over the $1\,000$ mock realizations, the posterior is well centered on the fiducial values, although residual shifts away from the exact fiducial point arise from parameter degeneracies (e.g., between the mass-richness relation scatter and the cosmological parameters). Nonetheless, we still recover all fiducials within the $1\sigma$ credible region, with all posterior-mean offsets $< 0.40\sigma$ from the fiducials. The maximum-posterior sample of the chain recovers all eight input parameters to better than $0.08\sigma$.

This test confirms statistically consistent recovery of the input values when the modeling choices for the inference match the assumptions behind the mock data. It also demonstrates that parameter updates propagate correctly through each stage at every sampling step, and that the modular pieces of the pipeline are correctly assembled into a single, functioning workflow.

\section{Application to the LSST-DESC cosmoDC2 simulation}
\label{sec:results}

To validate the \pipeline, we perform an end-to-end cosmological analysis of the \redmapper~cluster catalog derived from the LSST-DESC cosmoDC2 simulation. The analysis combines binned cluster abundance with the stacked excess surface density profile ($\Delta\Sigma$), using richness $\lambda$ as the cluster observable and the mass-richness relation (Eq.~\eqref{mass_richness_relation}), to jointly constrain two cosmological parameters, $\Omega_c$ and $\sigma_8$ from $\Lambda$CDM, together with the six associated mass-richness relation parameters.

\subsection{Dataset}
\label{dataset}

The LSST DESC \textit{Data Challenge~2} (DC2, \citealt{lsst2021lsst}) provides a suite of cosmological and image simulations designed to test and refine the analysis tools that will be used for LSST. From this suite, we use the cosmoDC2 extragalactic catalog.

\subsubsection{The cosmoDC2 extragalactic catalog}
\label{subsec:cosmodc2}

The cosmoDC2 simulation\footnote{\url{https://irsa.ipac.caltech.edu/Missions/cosmodc2.html}}~\citep{korytov2019cosmodc2} is a large synthetic galaxy catalog covering $\Omega_{\mathrm{DC2}} = 440~\mathrm{deg}^2$ and built on the \textit{Outer Rim} $N$-body simulation~\citep{Heitmann2019Outerim}; dark matter halos were identified via a Friends-of-Friends (FoF) algorithm and assigned both FoF masses ($M_{\mathrm{FoF}}$) and spherical overdensity masses ($M_{200c}$). Galaxies are then painted on halos using GalSampler~\citep{Hearin2020GalSampler} based on the Galacticus semi-analytic model~\citep{Benson2012}. cosmoDC2 compiles "truth" galaxy properties (along with host dark matter halo position, shape, and masses) such as stellar mass, intrinsic ellipticities, and local shear/convergence evaluation (obtained with ray-tracing), and broadband filter "true" magnitude to a redshift $z \sim 3$ and $r \simeq 28$. It has been extensively tested to fulfill science requirements for LSST preparation (see e.g. \citealt{Kovacs2022cosmoDC2}), including cluster properties. 

\subsubsection{\redmapper galaxy cluster catalog}

\redmapper is a cluster finder algorithm that identifies clusters through overdensities of red-sequence galaxies~\citep{Rykoff2014redmapper}. It has been applied to the cosmoDC2 extragalactic catalog, and each cosmoDC2 \redmapper detection is assigned a sky position, photometric redshift (computed from the photometric bands of identified member galaxies), and a richness\footnote{Computed as the sum of galaxy membership probabilities within a specific cluster aperture.} $\lambda$. The full cosmoDC2-\redmapper cluster catalog contains $\sim 880,000$ detections\footnote{Most of these are low-richness detections, abundant in the deep and noiseless cosmoDC2 photometry, where the catalog is impure.} within $5 < \lambda < 270$ and $0.1<z<1.15$. 

To assess the selection systematics of the cluster catalog, we match \redmapper-detected clusters to the latent cosmoDC2 halo population using a membership-based approach\footnote{\url{https://github.com/LSSTDESC/clevar}}.
We use the matched cluster-halo catalog to establish (i) the fiducial mass-richness relation in Eq.~\eqref{mass_richness_relation} (considering $M_\mathrm{piv}=10^{14.3}M_\odot$ and $z_\mathrm{piv}=0.5$) and (ii) the calibration of the \redmapper selection function. Let us note that the halo-to-cluster distance\footnote{The halo center in cosmoDC2 is the location of the FoF halo’s minimum gravitational potential~\citep{Heitmann2019Outerim}, where the central galaxy is placed~\citep{korytov2019cosmodc2}.} accessible through this matching procedure is essential to assess miscentering in weak lensing profile prediction. The matching procedure, fiducial scaling relation, selection function, and miscentering systematic studies are detailed in App.~\ref{appA}.

\subsection{Galaxy cluster binned probes}
\subsubsection{Measured abundance and stacked lensing profiles}
\label{cosmodc2_data_vectors}

\begin{figure*}[tbp]
    \centering
    \includegraphics[width=0.9\textwidth]{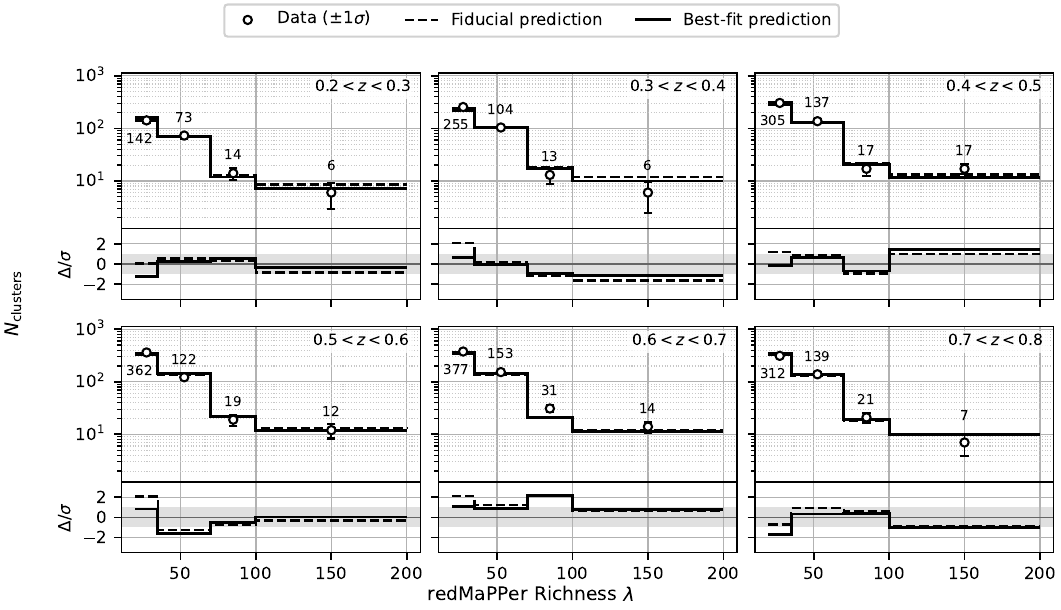}
    \caption{Cluster abundance as a function of richness $\lambda$ in redshift
    bins for the baseline \redmapper cosmoDC2 catalog. Each panel shows a
    redshift slice with (i) the measured cluster counts from \texttt{TXPipe} (points);
    (ii) the fiducial prediction (dashed step lines), i.e.\ the \texttt{CROW} prediction evaluated
    at the true cosmoDC2 cosmology and the fiducial \redmapper scaling-relation
    and selection-function parameters (App.~\ref{appA}); and (iii) the baseline
    fit (solid step lines), i.e.\ the same prediction with the eight sampled parameters set to the
    maximum posterior parameter values of the joint $N+\Delta\Sigma$ baseline analysis
    (Sec.~\ref{baseline_inference}), all non-sampled quantities kept at their
    fiducial values. Lower panels show the residuals $(N - N_{\rm pred})/\sigma$ for the fiducial (dashed) and baseline fit (solid) predictions, and the grey band marks $\pm1\sigma$.
    Data errors are computed with the theoretical covariance described in
    Sec.~\ref{baseline_config}.}
\label{fig:counts_richness}
\end{figure*}
\begin{figure*}[tbp]
\centering
\includegraphics[width=0.9\textwidth]{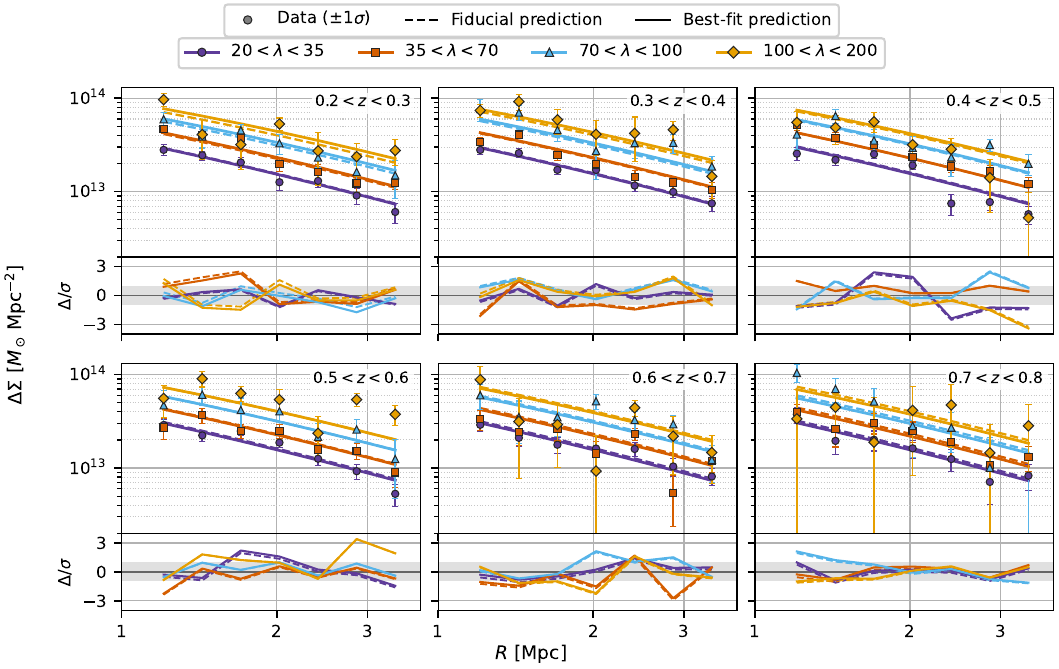}
\caption{Stacked excess surface density profiles for the baseline
\redmapper cosmoDC2 catalog. Each panel shows a redshift slice with (i)
the measured stacked lensing profiles from \texttt{TXPipe}  (points); (ii) the
fiducial prediction (dashed lines), as defined in Fig.~\ref{fig:counts_richness}; and
(iii) the baseline fit (solid lines), i.e.\ the prediction evaluated at the maximum posterior parameter values of the same joint fit, as defined in Fig.~\ref{fig:counts_richness}. Lower panels show the residuals $(\Delta\Sigma - \Delta\Sigma_{\rm pred})/\sigma$ for each richness bin, with the same line styles, and the grey band marks $\pm1\sigma$.
Error bars come from the bootstrap covariance of
Sec.~\ref{baseline_config}.}
\label{fig:delta_sigma}
\end{figure*}

We measure the cluster abundance and stacked weak lensing profiles using the \texttt{CLCluster} stages (Sec.~\ref{sec:txpipe}), configured for this analysis as follows.

We adopt the binning scheme in redshift-richness following the edges $z \in \{0.2, 0.3, 0.4, 0.5, 0.6, 0.7, 0.8\}$ and $\lambda \in \{20, 35, 70, 100, 200\}$, leaving 2658 clusters. The binning scheme is inspired from standard binned lensing analyses such as DES \citep{desclustercosmo20, abbott2025dark} and KiDS \citep{lesci2022amico}, and accounts for the degradation of the \redmapper  performance at low\footnote{Due to lack of $u$-band data in DES and KiDS, but will be seriously improved for LSST.} and high redshifts~\citep{McClintock2019mass, cui2025redmapper}, while the richness threshold  $\lambda \geq 20$ mimics the high-purity cuts generally used to ensure clean cluster samples. The resulting cluster abundance is shown in Fig.~\ref{fig:counts_richness}. 

For the source selection, we use the \texttt{TXPipe} ingestion and source-selection stages (\texttt{TXCosmoDC2Mock}, \texttt{TXSourceTomography} and \texttt{TXSourceSelectorMetadetect}) with a $\text{mag-}i \leq 24.25$ cut (truth cosmoDC2 photometry) and a global redshift cut $0.1 < z_\mathrm{true, gal} < 3$. The resulting source sample yields a mean source density of $n_\mathrm{gal} \approx 25\ \mathrm{arcmin}^{-2}$, consistent with 10 years of observations with LSST~\citep{mandelbaum2018lsst, Chang2013}. Next, for each \redmapper detection with assigned redshift $z_\mathrm{cl}$, galaxies are selected according to $z_\mathrm{true, gal} > z_\mathrm{cl} + 0.2$, identifying galaxies behind the cluster\footnote{The $\Delta z=0.2$ is common to mitigate projection effects and foreground contamination, which in the case of true redshifts is absent.}. 
Lensed ellipticities are computed by combining the local shear/convergence evaluation and an intrinsic ellipticity, taken to be a Gaussian variable with shape noise $\sigma_\mathrm{SN} = 0.26$ per component, matching observational estimates~\citep{Chang2013}.

We configure \texttt{TXPipe} to bin the source galaxies within seven logarithmically spaced radial bins within $1.1\ \mathrm{Mpc} < R < 3.7\ \mathrm{Mpc}$. The lower bound avoids signal attenuation observed at small scales in cosmoDC2 \citep{korytov2019cosmodc2, Kovacs2022cosmoDC2, Payerne2025dc2CL}, while the upper bound restricts to the one-halo regime. We show in Fig.~\ref{fig:delta_sigma} the measured $\Delta \Sigma$ stacked lensing profiles in the different richness-redshift bins. 
In App.~\ref{app:comparion_payerne2025}, we also compare the measured \redmapper cosmoDC2 stacked lensing profiles to the measurements from \citet{Payerne2025dc2CL} and find good agreement.

\subsubsection{Comparison to fiducial cosmological prediction}
\label{baseline_config}

We discuss the baseline modeling choices we use in this paper and compare data to our fiducial predictions. 

For both count and lensing predictions in Eqs.~\eqref{cluster_abundance} and~\eqref{cluster_theta_profile}, we use the \citet{despali2016universality} halo mass function prescription (set according to the $M_{200c}$ overdensity mass definition). For the scaling relation, we use the \cite{Murata2019} mass-richness relation (Eq.~\eqref{mass_richness_relation}). The selection function follows the completeness parametrization of \citet{Aguena2018} ($\Phi(M, x, z) = c(M, z)$)\footnote{We consider purity equal to one as discussed in Appendix~\ref{appA}}, with fiducial parameters obtained from the matching between cosmoDC2 halos and \redmapper detections (see Table~\ref{tab:selection}). We do not include projection effects in the baseline modeling.

For lensing, we model $\Delta\Sigma$ with the one-halo term using a Navarro–Frenk–White (NFW, \citealt{nfw}) profile with concentration $c_{\rm 200c}$=3.8, representative of the mean concentration observed in the cluster sample\footnote{This value is obtained by computing the mean concentration prediction with \citet{bhattacharya2013} c-M relation over the \redmapper sample, knowing the true dark matter halo mass from the paired cosmoDC2 halo.}. We fix the cluster concentration to a fiducial mean value for simplicity. Tests with a free concentration parameter showed negligible impact on the inferred constraints. Given our radial range ($R < 3.7$ Mpc), we model only the one-halo term in Eq.~\eqref{cluster_theta_profile}. We do not consider boost factor corrections, since member contamination in cosmoDC2 is below 0.1\% even for photo-$z$-selected sources~\citep{Payerne2025dc2CL}. We show in App.~\ref{appA} that miscentering\footnote{Offset between the cluster-finder center and the underlying halo's center of mass.} has negligible impact on the modeling, due to (i) the low fraction of miscentered clusters and (ii) the low radial cut at $R > 1$ Mpc we consider in this work. In the same Appendix, we show that the scatter in the \redmapper cluster photometric redshifts also has a low impact on the modeling.

The theoretical cluster count covariance matrix (Eq.~\eqref{ssc_cov}) combines a Poisson shot-noise term with the SSC contribution, assuming the \citet{tinker2010large} halo bias prescription. This covariance is evaluated at the fiducial cosmoDC2 cosmology and held fixed throughout inference (Sec.~\ref{sec:firecrown}). For weak lensing, we adopt a bootstrap covariance with $100$ resamples in each redshift-richness bin of the \redmapper cluster lensing profiles (we neglect off-diagonal covariance terms since they are very noisy). Between different richness-redshift bins, the covariance between stacked profiles and between different radial scales is generally ignored, as it is significantly smaller than the per-bin components. 

We compare in Fig.~\ref{fig:counts_richness} and Fig.~\ref{fig:delta_sigma} the fiducial and best-fit predictions to the data vectors, with their residuals in the lower panels. Most residuals lie within $2\sigma$, which shows that the baseline model reproduces the overall amplitude and shape of both data vectors. The lensing error bars can be improved with a theoretical covariance matrix, which is in development (Sec.~\ref{sec:discussion}).

\subsection{Cosmological inference}
\label{cosmo_constraints}
We sample the joint ($N+\Delta\Sigma$) posterior distribution of an eight-dimensional parameter space ($\mu_0, \mu_m, \mu_z, \sigma_0, \sigma_m, \sigma_z$, $\Omega_c$, $\sigma_8$) with \texttt{emcee} within the \texttt{CosmoSIS} framework, using the Gaussian likelihoods described in Sec.~\ref{sec:firecrown}\footnote{Although our highest-richness bins contain few clusters, \citet{Payerne2024} showed that the Gaussian likelihood recovers parameter errors to within $5\%$ even for bins populated by fewer than five clusters, while a Poisson description underestimates them by $\sim30\%$ as it cannot account for the sample variance.}. We adopt flat priors and fiducials described in Table~\ref{tab:fiducial}. Each chain contains over $2.2 \times 10^6$ samples generated with $100$ walkers, with the first 15\% discarded as burn-in. We present the baseline analysis in Sec.~\ref{baseline_inference}, and, in Sec.~\ref{modeling_choices}, vary one ingredient of the analysis at a time (the halo mass function and the fiducial cosmology input) to demonstrate the pipeline's flexibility.

\begin{figure*}[p]
\centering\includegraphics[width=1.0\textwidth]{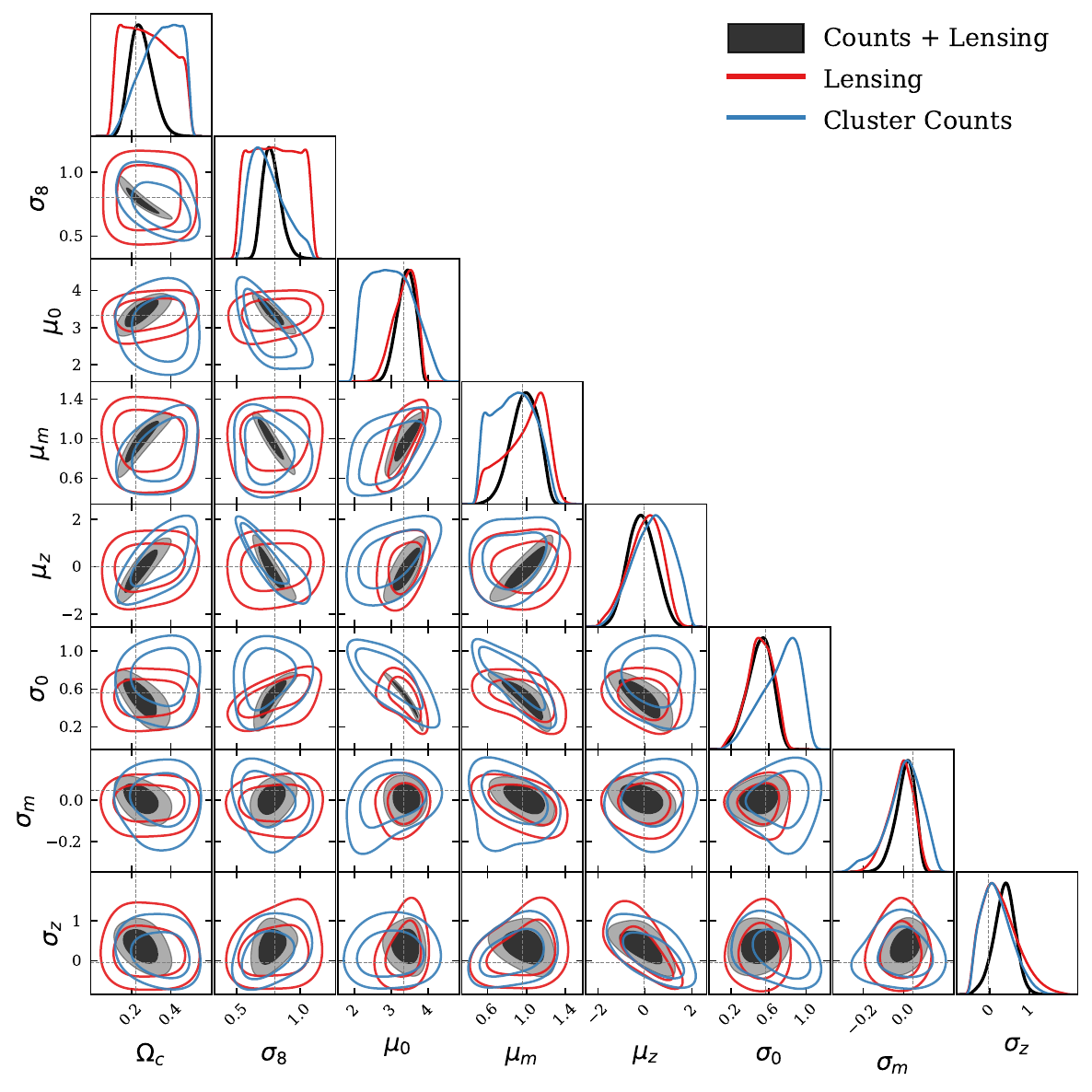}
    \caption{Posterior constraints on the six mass--richness relation parameters and the cosmological parameters $\Omega_c$ and $\sigma_8$ for the cosmoDC2 \redmapper cluster dataset. Contours show the 68\% and 95\% credible regions obtained from cluster abundance alone (blue), stacked weak lensing alone (red), and their combination (black). Dashed lines indicate the fiducial cosmoDC2 cosmology and the fiducial mass--richness parameters from Table~\ref{tab:fiducial}.}
    \label{fig:full_fit_baseline}
\end{figure*}

\subsubsection{\redmapper CC+WL: Baseline analysis}
\label{baseline_inference}
Figure~\ref{fig:full_fit_baseline} shows the posterior constraints obtained from cluster abundance alone, lensing alone, and their combination. For the cosmological parameters, the lensing-only fit remains prior-dominated, while cluster counts alone provide slightly tighter constraints on $\Omega_c$ and $\sigma_8$. The overlap between contours and the different degeneracy directions between count-only and lensing-only constraints (especially in the $\sigma_z-[\mu_0, \mu_m, \mu_z]$ plane) make their combination powerful to significantly increase parameter precision. The complementarity between probes is especially visible in the $\sigma_8$--$\Omega_c$ plane, where posteriors break free from the prior-dominated square contours. For the joint case, the fiducial cosmoDC2 cosmology is recovered well within $1\sigma$, with posterior means offset by only $0.51\sigma$ in $\Omega_c$ and $0.34\sigma$ in $\sigma_8$ from the fiducial values. The corresponding baseline-fit (maximum posterior values) predictions are over-plotted on data vectors in Figs.~\ref{fig:counts_richness} and~\ref{fig:delta_sigma}.

We also compare in App.~\ref{app:comparion_payerne2025} our fixed-cosmology results (sampling only the posterior of the cluster scaling relation parameters) to the work of \citet{Payerne2025dc2CL} who inferred, independently of this work, the \redmapper cluster scaling relation from the cosmoDC2-\redmapper abundance and lensing measurements. The comparison shows overall good agreement despite differences in modeling choices and data processing.

\subsubsection{\redmapper CC+WL: Variations across different analysis setups}
\label{modeling_choices}

\begin{figure*}[!t]
\centering
\includegraphics[width=0.85\textwidth]{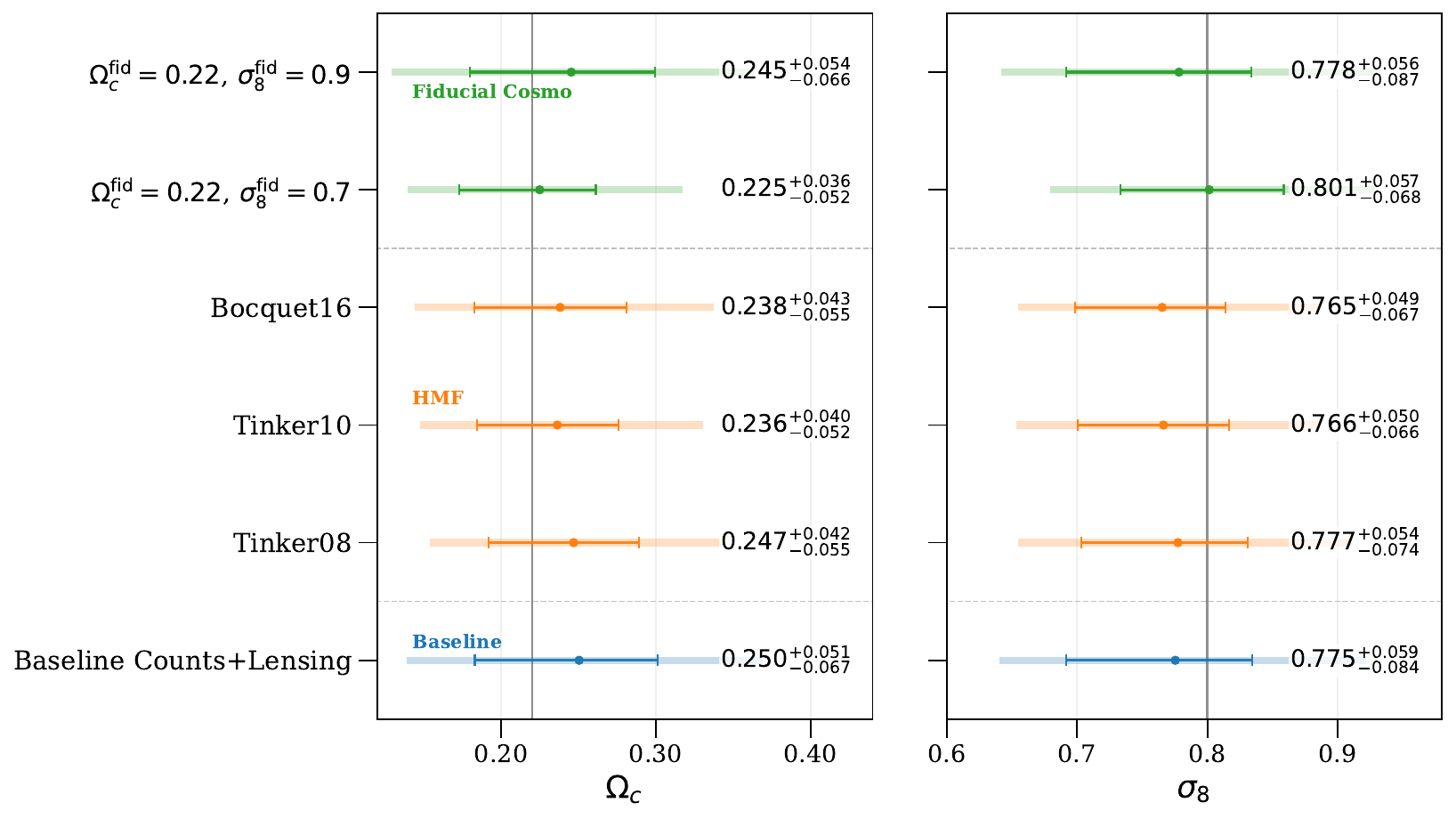}
\caption{Summary of the marginalized constraints on $\Omega_c$ (left) and
$\sigma_8$ (right) for the baseline analysis and the validation tests
described in Sec.~\ref{modeling_choices}. For each case, the point marks
the marginalized posterior mean and the dark (light) bars the $1\sigma$
($2\sigma$) intervals, both derived from the posterior chains; the
corresponding mean $\pm\,1\sigma$ values are annotated on each row.
Vertical lines indicate the fiducial cosmoDC2 values, and colors group
the different analysis variations. All cases recover the fiducial
cosmology within $1\sigma$.}
\label{summary}
\end{figure*}

The \pipeline's modular design allows individual ingredients of the analysis, from the halo mass function to the fiducial cosmology, to be swapped without changing the rest of the pipeline. We illustrate this flexibility here by testing the robustness of our fiducial cosmoDC2 constraints against alternative analysis configurations, summarized in Fig.~\ref{summary}.\\

\noindent\textit{Varying fiducial cosmological parameters:} 
Estimating the stacked weak lensing profiles and the theoretical cluster counts covariance requires a priori knowledge of the fiducial background cosmology. We show in App.~\ref{app:ds_profiles_diff_fiducial} the ratio between the lensing estimators computed at the fiducial cosmology (taking $\Omega_c=0.22$) and for different $\Omega_c=\{0.20, 0.24\}$. Varying $\Omega_c$ can shift the lensing profiles up to $\sim 5\%$. This dependence can be accounted for by applying cosmology-dependent corrections to the lensing data vector~\citep{sereno2015comalit, desclustercosmo20, lesci2022amico}. Since we analyze a simulation with known input cosmology, we do not apply these corrections and leave their implementation to the pipeline for future work.

Moreover, the cluster count covariance in Eq.~\eqref{ssc_cov} is computed once at the fiducial cosmoDC2 cosmology. In principle, the covariance matrix should be updated at each MCMC evaluation, and the consequences of keeping it fixed have already been explored~\citep{fumagalli2021euclid, Lettieri2026covariance}. To test the impact of this choice, we update our fiducial $\sigma_8$ value and perform our baseline parameter estimation, which leaves the computation of the lensing estimators unaffected. Changing the fiducial cosmology only requires editing the cosmological parameter values in a single shared configuration file, which is then propagated through every stage of the pipeline.

We vary $\sigma_8 \in \{0.70, 0.90\}$ and estimate the corresponding parameter posterior distributions. Fig.~\ref{summary} shows a summary of the resulting cosmological constraints. Relative to the baseline cosmology, the $\sigma_8=0.70$ case produces mean posterior shifts of $0.34\sigma$ and $0.27\sigma$ for $\Omega_c$ and $\sigma_8$, respectively, while the shifts are only $0.06\sigma$ and $0.03\sigma$ for $\sigma_8=0.90$. Changes in the fiducial cosmology also modify the covariance and are therefore expected to affect both the posterior means and constraining power. Despite these variations, the recovered cosmological parameters remain consistent with the true cosmoDC2 values within $1\sigma$ in all cases. \\

\noindent\textit{Varying the halo mass function}: In our analysis, we use \citet{despali2016universality} halo mass function (HMF). Several other HMFs \citep{Tinker2008,tinker2010large,Bocquet2019}, calibrated on simulations, are available in the cluster literature to predict the underlying halo mass-redshift distribution in the Universe, and can differ from \citet{despali2016universality} prediction for a variety of halo mass and redshift ranges\footnote{For the fiducial cosmoDC2 cosmology, the predicted halo abundances from these HMFs differ from Despali16 at the few-to-$\sim20\%$ level.}. We want to test the impact of using different HMFs on parameter constraints, namely \citet{Tinker2008}, \citet{tinker2010large} and \citet{Bocquet2019}.
In our pipeline, switching between them requires editing a single entry in the \texttt{CLPFirecrown} configuration file (the \texttt{hmf} field in Listing~\ref{lst:firecrown_config}), with no changes to the pipeline code.

Fig.~\ref{summary} shows the results as we perform the inference for different HMFs. Although the HMF choice affects our predictions, part of this effect is reduced by degeneracies with the mass--richness relation parameters. As a result, the different HMFs produce similar posterior contours, and we recover the true cosmoDC2 cosmology within $1\sigma$ in all cases. \citet{Tinker2008} shows the best agreement with the baseline, with less than 0.05$\sigma$ shifts to the mean posterior on both parameters. The largest shift in $\Omega_c$ is obtained for \citet{tinker2010large} ($\sim0.19\sigma$ from the baseline), and \citet{Bocquet2019} for $\sigma_8$ ($\sim 0.11\sigma$ from the baseline).

\section{Discussion and perspectives}
\label{sec:discussion}

In this work, we present the first version of the fully integrated \pipeline for analyzing LSST cluster data. In its current version, it enables cosmological analyses starting from galaxy and cluster catalogs to parameter posteriors and relies on several tools developed within the collaboration: \texttt{Ceci} for pipeline integration, \texttt{TXPipe} for data ingestion and formatting, \texttt{CCL}, \texttt{CROW}, and \texttt{CLMM} for modeling cluster probes, \texttt{TJPCov} for theoretical covariance computation, and \texttt{Firecrown} for Bayesian inference.

This pipeline represents a significant collaborative achievement, developed jointly by members of the DESC cluster working group and the broader LSST-DESC collaboration. The pipeline is modular by construction, such that the data measurement, theoretical modeling, covariance, and inference stages can each be modified or replaced separately, allowing us to adapt to different analysis choices. \texttt{Ceci} continuously integrates these stages into a single coherent pipeline, and cosmological parameters are propagated consistently across all stages through a shared \texttt{CCL} backend. Major modeling choices can be changed by editing a single string in a configuration file. New data formats can similarly be accommodated by updating the relevant input entry points. Finally, the pipeline is built to run on batch computing farms, with \texttt{Ceci}-managed MPI parallelization distributing computationally intensive stages across multiple processes to reduce wall-clock time and memory usage.

We first tested the inference part of the pipeline with a mean mock dataset well described by our probe predictions, and found statistically consistent recovery of the fiducial parameters. We then performed a first full run of the pipeline by analyzing the binned abundance and $\Delta\Sigma$ lensing profile of galaxy clusters detected with \redmapper on the cosmoDC2 truth galaxy catalog. Using cosmoDC2 allows us to test the pipeline in ideal conditions, as the true galaxy properties are known, especially shapes and redshifts, with no survey systematics. The outcome of the pipeline run is the posterior distribution of the cosmological parameters and of the \redmapper cluster scaling relation parameters, which showed good overall agreement with the fiducial values of the simulated dataset. We also showed the modularity of the analysis for different configurations (different fiducial cosmology and HMF prescription), recovering fiducials within the credible region in all cases. 

Beyond this idealized setup, which constitutes a first-step validation of the workflow, further work is needed for analyses with real cluster and lensing datasets. We discuss below the observational challenges of moving to real LSST data, what remains specific to the binned $N+\Delta\Sigma$ baseline validated here, how the pipeline extends beyond this baseline, and, finally, planned improvements to covariance and inference.\\

\noindent \textit{Observational systematics.}
Real LSST data introduces several systematics relevant for cluster cosmology. In our cosmoDC2 analysis, clusters' background galaxies are selected via their true redshift, which is also used to compute the lensing-source geometrical weights. LSST galaxy redshifts, by contrast, will be estimated from noisy $ugrizy$ photometry~\citep{zhang2026photometric}, potentially biasing the lensing profile through outliers\footnote{True foreground galaxies identified as background galaxies.} and cluster members\footnote{Galaxies physically associated with the cluster but selected as background sources, which can also contaminate the signal through their intrinsic alignments~\citep{blazek2012separating, chisari2014intrinsic}.} entering the source sample, and through biased lensing-source distance weights that cannot be mitigated simply by increasing sample size~\citep{Simet2017SDSSmassrichness, Applegate2014}. Although unused in our test case, this pipeline is already capable of handling photometric redshift inputs in \texttt{TXPipe} for data vector construction, but is under development on the modeling side. Additionally, a tomographic bin-based source selection is already under development in \texttt{TXPipe} as an alternative, replacing individual point estimates with source-redshift distributions per tomographic bin.

Shear calibration is not exercised in the cosmoDC2 idealized test, since true galaxy shapes are directly provided. For LSST, shapes will instead be measured from images, where PSF modeling~\citep{arbelaez2026deep}, blending~\citep{ramel2026probabilisticcharacterizationblendinglsst}, and the calibration method itself~\citep{metacal, zhou2026accurate} introduce complex, potentially non-linear relationships between observed and true shapes. Currently, DESC plans to provide different shear catalogs for LSST, such as \texttt{Metadetect}~\citep{sheldon2020metadetect, sheldon2023metadetect} and \texttt{AnaCal}~\citep{li2023analytical, li2024differentiable}. The \texttt{TXPipe} component of the \pipeline is already adapted to use \texttt{Metadetect} and is being extended to \texttt{AnaCal} catalogs. 

Finally, cluster finders will rely on realistic LSST multi-band photometry and survey properties (PSF, depth variations, blending, etc.), which can introduce several sources of systematic bias in the cluster catalogs (miscentering,~\citealt{ding2025miscentring}; uncertain cluster photometric redshift estimates, selection function biases,~\citealt{Aguena2018}; etc.). The selection function is modeled within \texttt{CROW} as a standalone, modular object, so further implementations can be easily incorporated. Projection effects are also being added to the theoretical predictions. The richness–shear covariance induced by the cluster selection may also need to be modeled~\citep{zhang2024impact}.\\

\noindent \textit{Refining the binned counts and stacked $\Delta\Sigma$ baseline.} 
Our current lensing baseline deliberately includes only the one-halo regime, excluding scales where miscentering, member contamination~\citep{Gruen2015covDS, McClintock2019mass}, and neighboring-cluster contributions become important. At small scales ($\sim 1$ Mpc), where the lensing amplitude is strongest, these effects tend to bias or dilute the signal, while at larger scales the two-halo contribution and the fiducial-cosmology dependence of $\Delta\Sigma$ both grow~\citep{sereno2015comalit}. \texttt{CROW} already supports a two-halo term and individual shear profile miscentering corrections via \texttt{CLMM}, as configurable modeling choices, although the current implementation to the binned theoretical predictions is computationally slow and must be optimized. Extending our radial range requires validating and optimizing these existing capabilities and further modeling development to incorporate refined halo concentration modeling, cluster triaxiality~\citep{fu2026constraining}, halo orientation~\citep{Dietrich2014}, and the fiducial cosmology correction to the shear profile predictions~\citep{sereno2015comalit}.\\

\noindent \textit{Beyond this baseline.}
The \pipeline is designed to support analysis choices beyond the binned $N+\Delta\Sigma$ combination presented here. The pipeline can also run end-to-end with the reduced tangential shear $g_t$ profile instead of $\Delta\Sigma$. Individual, unbinned cluster lensing profiles are already computed and saved in \texttt{TXPipe}, but the complete pipeline workflow is not yet fully developed for an unbinned cluster analysis. Combining cluster probes with other LSST-DESC probes in a joint $6\times2$pt$+N$ analysis requires additional pipeline stages and further development at the likelihood level. This development should be easier for probe combinations already implemented within the same infrastructure, such as the standard LSST-DESC $3\times2$pt analysis (weak lensing and galaxy clustering) that uses the same DESC tools~\citep{Txpipe}.\\

\noindent \textit{Covariance and inference.}
The cluster count covariance used in this work is evaluated once, at a single fiducial cosmology, and held fixed throughout inference. In principle, it should instead be recomputed at each parameter configuration explored by the sampler. Additionally, a theoretical covariance model for cluster weak lensing, beyond the data-driven bootstrap covariance used in this work, is also still to be implemented and will be needed to capture contributions such as intrinsic profile scatter and correlated large-scale structure~\citep{Wu2019covDeltaSigma}.
Given the pipeline's modular design, it can also be extended to support Simulation-Based Inference (SBI) as an alternative inference approach~\citep{payerne2026simulation}. This could be achieved by introducing an additional inference stage while leaving the data measurement and theoretical modeling stages described here unchanged, with the theory modules providing the forward simulator required by the SBI framework.
 
\subsection*{Data and Software availability}
The data products required to reproduce the results presented in this work,
including the \texttt{SACC} data vectors, the posterior chains, and the mock realizations, are hosted on the NERSC Science Gateway
portal at \url{https://portal.nersc.gov/cfs/lsst/clpipe/data/}.
Regarding the DESC tools, the results of this work were obtained with \texttt{CLPipe} v1.0.0, which runs \texttt{TXPipe} (tag \texttt{clpipe-paper-v1}), \texttt{CROW} v1.0.12, \texttt{TJPCov} v0.6.0 and \texttt{Firecrown} v1.15.0a1.dev187 (commit \texttt{7098afc}). These codes rely on \texttt{CCL} v3.2.1, \texttt{CLMM} v1.16.1 and \texttt{SACC} v2.1.2. The stages are connected with \texttt{Ceci} v2.5. All DESC codes are publicly available at \url{https://github.com/LSSTDESC}.

\subsection*{Acknowledgments}


\textit{General acknowledgments.} EJB acknowledges support from the Enigmass+ research federation (CNRS, Universit\'e Grenoble Alpes, Universit\'e Savoie Mont-Blanc). CA was supported by DOE grant DE-SC009193 and the Leinweber Foundation. CGG was supported by the C\'esar Nombela  Research Talent Attraction grant from the Community of Madrid  (Ref. 2025-T1/TEC-36302). This work is supported by the French National Research Agency in the framework of the ``France 2030'' (ANR-15-IDEX-02).
The DESC acknowledges ongoing support from the Institut National de 
Physique Nucl\'eaire et de Physique des Particules in France; the 
Science \& Technology Facilities Council in the United Kingdom; and the
Department of Energy and the LSST Discovery Alliance
in the United States.  DESC uses resources of the IN2P3 
Computing Center (CC-IN2P3--Lyon/Villeurbanne - France) funded by the 
Centre National de la Recherche Scientifique; the National Energy 
Research Scientific Computing Center, a DOE Office of Science User 
Facility supported by the Office of Science of the U.S.\ Department of
Energy under Contract No.\ DE-AC02-05CH11231; STFC DiRAC HPC Facilities, 
funded by UK BEIS National E-infrastructure capital grants; and the UK 
particle physics grid, supported by the GridPP Collaboration.  This 
work was performed in part under DOE Contract DE-AC02-76SF00515.

This paper has undergone internal review in the LSST Dark Energy Science Collaboration. We are grateful to the internal reviewers, Ian Dell'Antonio and Rance Solomon, who provided helpful feedback and suggestions on the manuscript. We thank Matthew R. Becker for his work on the cluster prediction module of \texttt{Firecrown}, which later became \texttt{CROW}.

\textit{Author contributions.} EJB: Leading the analysis, development lead of \texttt{CLPipe}, co-lead developer of \texttt{CROW}, contributions to \texttt{TXPipe}, \texttt{Firecrown}, \texttt{TJPCov} and \texttt{SACC}, original draft and manuscript writing.
CP: Conceptualization, methodology, \clmm\ development, DC2-like mocks with \texttt{capish} and \clmm, data analysis, manuscript writing and review.
MAg: \texttt{CROW} co-developer, co-lead of \clmm\ development, review of the early draft.
CC: \clmm\ development, initial cluster weak lensing measurement in \texttt{TXPipe}, early definition of the \texttt{SACC} cluster tracers, pipeline design discussions, extensive feedback on the original draft.
SE: \texttt{CLPipe} development, analysis templates and code organization, installation of the conda environments for \texttt{Firecrown} and \texttt{TXPipe} that enabled the analysis.
TG: Project conceptualization and management, discussions of the analysis steps, help with the analysis of the results, review of the draft.
MR: DESC Builder, Cluster Working Group convener, cluster modules in \texttt{TXPipe} and use of \texttt{ceci}, co-lead of \clmm\ development, review of the early draft.
SDPV: \texttt{Firecrown} development, initial cluster counts framework, numerical integration backend, cross-checks with \texttt{NumCosmo}.
JZ: \texttt{TXPipe} maintainer, cluster modules in \texttt{TXPipe}, help with the analysis, review of the manuscript.
NA: Scientific discussions, debugging, interpretation of the results, manuscript feedback.
MAp: Contributions to \texttt{Firecrown} and design of the cluster counts framework.
CA: Early development of the cluster code in \texttt{TXPipe}.
DB: DESC Builder, precursor studies, scientific discussions, early testing and debugging.
NEC: DESC Builder, \ccl\ development, comments on the final draft.
AF: Implementation of the super-sample covariance in \texttt{TJPCov}.
CGG: Review of the cluster covariance implementation in \texttt{TJPCov}.
NLM: Scientific discussions, testing and debugging, interpretation of the results, review of the draft.
AIM: DESC Builder, early contributions to \texttt{RAIL} and \clmm\ conceptualization, funding acquisition, project administration, software and validation.
MP: \texttt{Firecrown} development.
EMP: \texttt{TXPipe} maintainer.
MPL: \texttt{NumCosmo} halo and cluster codes used to validate \texttt{Firecrown}, design of the initial cluster counts code.
ESR: \texttt{redMaPPer} development, \texttt{redMaPPer} run on cosmoDC2.
N\v{S}: DESC Builder, \ccl\ and \texttt{TJPCov} development.
RS: Internal reviewer of this manuscript.
KT: Explanatory notebooks for the \texttt{TXPipe} analysis stages.
YZ: \texttt{Firecrown} development, initial cluster counts framework.

\bibliographystyle{mnras}
\bibliography{main}
\appendix

\section{cosmoDC2-\redmapper cluster catalog}
\label{appA}

\begin{figure*}[tbp]
    \centering
    \includegraphics[width=0.48\textwidth]{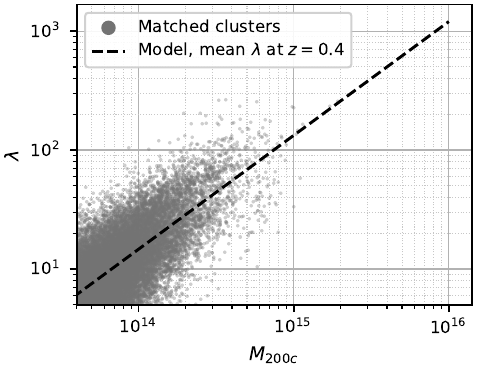}\hfill
    \includegraphics[width=0.48\textwidth]{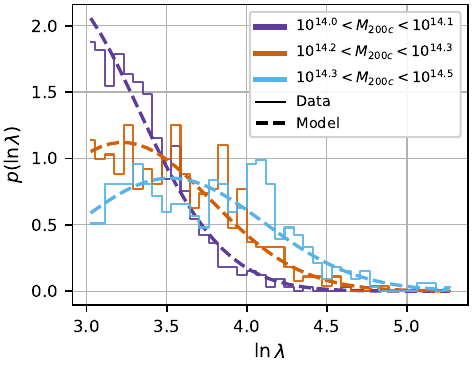}
    \caption{
    Validation of the fitted mass--richness relation. \textit{Left}: mass--richness scatter plot for the matched \redmapper cosmoDC2 catalog. The dashed line shows Eq.~\eqref{mass_richness_relation} evaluated at the best-fit parameter values for $z=0.4$. \textit{Right}: normalized distributions of $\ln\lambda$ in three halo mass bins (histograms) compared with the corresponding model predictions (dashed lines) obtained using the best-fit parameter values.
    }
    \label{fig:mass_richness}
\end{figure*}

To derive fiducial parameters and analyze cluster properties, we match detected clusters to their true halo counterparts using membership-based matching via the ClEvaR\footnote{\url{https://github.com/LSSTDESC/clevar}} software. The matching procedure is performed after applying cuts of
$M_{200c}>10^{13}M_\odot$ to the halo catalog and
$\lambda>5$ to the \redmapper cluster catalog. Each cross-matched cluster is then assigned a true spherical overdensity mass $M_{200c}$, a true $\text{RA}_{true}$, $\text{DEC}_{true}$, and a true redshift $z_{true}$ from its associated cosmoDC2 halo. 

\subsection{Mass-Richness}

With the matched catalog, we fit the six parameters of the mass-richness relation (Eq.~\eqref{mass_richness_relation}) using MCMC sampling with Gaussian likelihood and truncation correction~\citep{kelly2007some}. The fit uses pivot values $M_\mathrm{p} = 10^{14.3}\,M_\odot$ and $z_\mathrm{p} = 0.5$, with cuts $M_{200c} > 4\times10^{13}\,M_\odot
$, $10 < \lambda < 400$, $0.2 < z < 0.8$. The mass and richness cuts follow \citet{Payerne2025dc2CL}, who test the stability of this fiducial relation against these choices. The mass cut excludes low-mass halos that would bias the fit, while the redshift cut reflects the degraded red-sequence detection of \redmapper at higher redshifts~\citep{Rykoff2014redmapper}.

The quality of the fit is illustrated in Fig.~\ref{fig:mass_richness}.  Both panels demonstrate that the fitted mass--richness relation provides an accurate description of the observed richness distribution. The posterior mean values are listed in Table~\ref{tab:fiducial} along with the used priors.

\subsection{Cluster vs.\ Halo Redshift}
We compare the \redmapper photometric cluster redshifts to the true halo redshifts from the cosmoDC2 catalog for all matched pairs. The fractional redshift difference, $(z_\mathrm{cluster}-z_\mathrm{halo})/(1+z_\mathrm{halo})$, has a mean of $-3\times10^{-4}$ and a scatter of $0.0105$. The absence of a significant systematic bias indicates that no redshift correction is required, and we therefore adopt the \redmapper redshifts throughout the analysis.

\subsection{Selection Function}
\label{sec:selfunc}

For the baseline cosmoDC2 analysis, we model the selection function $\Phi$ in Eqs.~\eqref{cluster_abundance} and~\eqref{cluster_theta_profile}, following the prescription of \cite{Aguena2018}\footnote{In this work we make a small change where the purity model is computed on the $\ln \lambda$ scale.}, such that $\Phi(M, \lambda, z) = c(M, z)/p(\lambda, z)$, where we model completeness ($c(M, z)$) and purity ($p(\lambda, z)$) as
\begin{align}
    \label{selection1}
    c(M, z) &= \frac{N_{\mathrm{mat}}}{N_{\mathrm{halos}}}
    = \frac{\left(M/M_0\right)^{n_c}}
    {\left(M/M_0\right)^{n_c} + 1},
\end{align}
\begin{align}
\label{selection2}
    p(\lambda, z) &= \frac{N_{\mathrm{mat}}}{N_{\mathrm{clusters}}}
    = \frac{\left(\ln\lambda/\ln\lambda_0\right)^{n_p}}
    {\left(\ln\lambda/\ln\lambda_0\right)^{n_p} + 1}
,\end{align}
with
\begin{align}
\label{selection_params}
    n_{c} &= c_0 + c_1(1+z) \\
    n_{p} &= p_0 + p_1(1+z) \\
    \log M_0&= \log a_c + \log b_c (1 + z) \\
    \ln \lambda_0 &= \ln a_p + \ln b_p (1 + z)
,\end{align}
and $N_\mathrm{mat}$, $N_\mathrm{halos}$, and $N_\mathrm{clusters}$ denote the number of matched cluster--halo pairs, the total number of halos, and the total number of detected clusters, respectively.

\begin{figure*}[t]
    \centering\includegraphics[width=\textwidth]{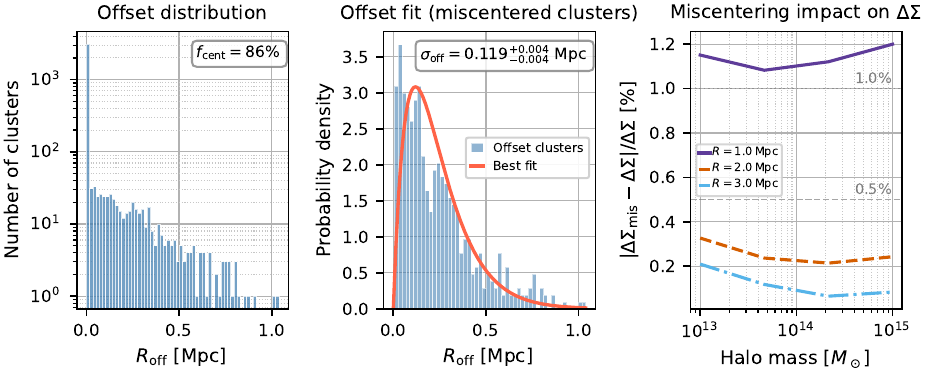}
    \caption{\textit{Left:} Distribution of projected offsets $R_\mathrm{off}$ between \redmapper cluster centers and their matched cosmoDC2 halo centers, on a logarithmic scale. \textit{Center:} Best-fit offset distribution (Eq.~\eqref{eq:p_roff}) for the non-zero offset subsample, given as $\sigma_\mathrm{off}=0.12\,\mathrm{Mpc}$. \textit{Right:} Relative difference between individual $\Delta\Sigma$ profiles computed with and without miscentering at $z=0.5$, using the measured offset fraction and best-fit $\sigma_\mathrm{off}$.}
\label{fig:miscentering}
\end{figure*}

A detailed characterization of the cosmoDC2 \redmapper cluster catalog is presented in Ricci et al., in prep. The fitted fiducial values are reported in Table~\ref{tab:selection}. For $\lambda>20$, the catalog is $\sim100\%$ pure and so we consider $p(\lambda, z) = 1$ above this richness threshold. Its completeness is approximately $20\%$, $70\%$, and $100\%$ for $10^{13} < M_{200c}/M_\odot < 10^{13.6}$, $10^{13.6} < M_{200c}/M_\odot < 10^{14.2}$, and higher masses, respectively.

\begin{table}[t]
    \centering
    \caption{Fiducial selection function parameters of the matched \redmapper--cosmoDC2     
    catalog, following the model of~\cite{Aguena2018} 
    (Eqs.~\eqref{selection1}--\eqref{selection_params}).}
    \begin{tabular}{lc}
        \hline\hline
        Parameter & Fiducial Value \\
        \hline
        $c_0$   & $1.13$          \\
        $c_1$    & $0.77$       \\
        $\log a_c$   & $13.31$    \\
        $\log b_c$   & $0.20$   \\
        $p_0$   & $1.98$        \\
        $p_1$    & $0.81$      \\
        $\ln a_p$   & $ 2.22$  \\
        $\ln b_p$   &  $-0.66$ \\
        \hline
    \end{tabular}
    \label{tab:selection}
\end{table}

\subsection{Miscentering}
\label{sec:miscentering}

We measure the projected offset $R_\mathrm{off}$ between each \redmapper cluster center and its matched halo center, assuming that the lensed positions of the central galaxies of the cosmoDC2 halos define the true cluster centers. We consider cluster--halo pairs with $M_{200c}\geq4\times10^{13}M_\odot$, $\lambda\geq20$, and $0.2\leq z\leq1.0$. The resulting offset distribution is shown in the left panel of Fig.~\ref{fig:miscentering}. Approximately 86\% of clusters are perfectly centered (i.e., \redmapper selects the true central galaxy, giving a null offset), while the full sample has a mean projected offset of $\langle R_\mathrm{off}\rangle = 0.033\,\mathrm{Mpc}$.

To characterize the non-zero offset population, we follow Appendix A of~\cite{Payerne2025dc2CL} and model the offset distribution with a Gamma distribution as \citep{giocoli2021amico,McClintock2019mass, zhang2019dark},
\begin{equation}
p(R_\mathrm{off} \mid \sigma_\mathrm{off}) =
\frac{R_\mathrm{off}}{\sigma_\mathrm{off}^2}
\exp\left(-\frac{R_\mathrm{off}}{\sigma_\mathrm{off}}\right),
\label{eq:p_roff}
\end{equation}
which gives a best-fit scale radius of $\sigma_\mathrm{off}=0.12\,\mathrm{Mpc}$ (center panel of Fig.~\ref{fig:miscentering}).

To assess the impact of miscentering on the weak-lensing signal, we use \texttt{CROW} to compute individual cluster $\Delta\Sigma$ profiles at $z=0.5$ with and without the miscentering correction, using the measured offset fraction and best-fit $\sigma_\mathrm{off}$. The resulting change in the predicted signal is only
1.2\%, 0.4\%, and 0.2\% at
$R=1$, $2$, and $3\,\mathrm{Mpc}$ respectively (right panel of Fig.~\ref{fig:miscentering}). We therefore neglect miscentering effects in this work.\\

\section{Stacked excess surface density profile measurement: impact of fiducial cosmology}
\label{app:ds_profiles_diff_fiducial}

\begin{figure}[h]
\centering
\includegraphics[width=\columnwidth]{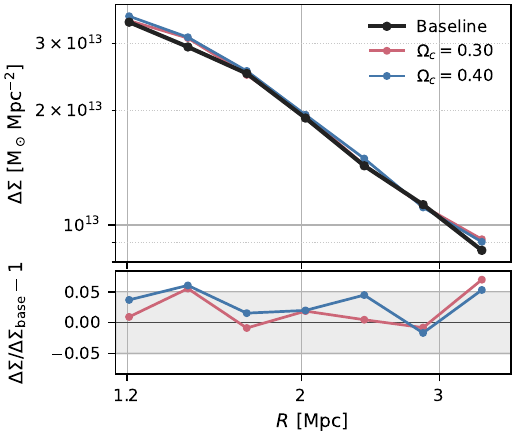}
\caption{Stacked weak lensing profiles estimated from data with different fiducial cosmologies to search for galaxy sources. The lower panel shows the fractional difference with respect to the baseline profile, and the grey band marks $\pm5\%$. The profiles were calculated on a single  redshift-richness bin.}
\label{fig:delta_cosmo_data}
\end{figure}

The estimation of the $\widehat{\Delta\Sigma}(R)$ profile depends on the choice of a fiducial cosmology, which enters the computation of the geometric lensing weights and the cluster--source galaxy associations. Variations in the fiducial cosmology therefore affect the measured data vector, with inferred weak-lensing masses changing by approximately $5$--$10\%$~\citep{cosmo_mass_bias}. This dependence can be accounted for at the modeling level during cosmological inference~\citep{sereno2015comalit}. To illustrate the impact of this effect in our analysis, we estimate the stacked weak-lensing profile from our baseline using a few different fiducial cosmologies. For simplicity, we combine all clusters in the redshift range $0.2 < z < 0.8$ and richness range $20 < \lambda < 200$ into a single stacked profile. Fig.~\ref{fig:delta_cosmo_data} shows that changing the fiducial cosmology modifies the measured $\widehat{\Delta\Sigma}(R)$ profile by $\sim 5\%$ relative to the baseline cosmology.

\begin{figure*}[t]
    \centering
    \includegraphics[width=0.7\textwidth]{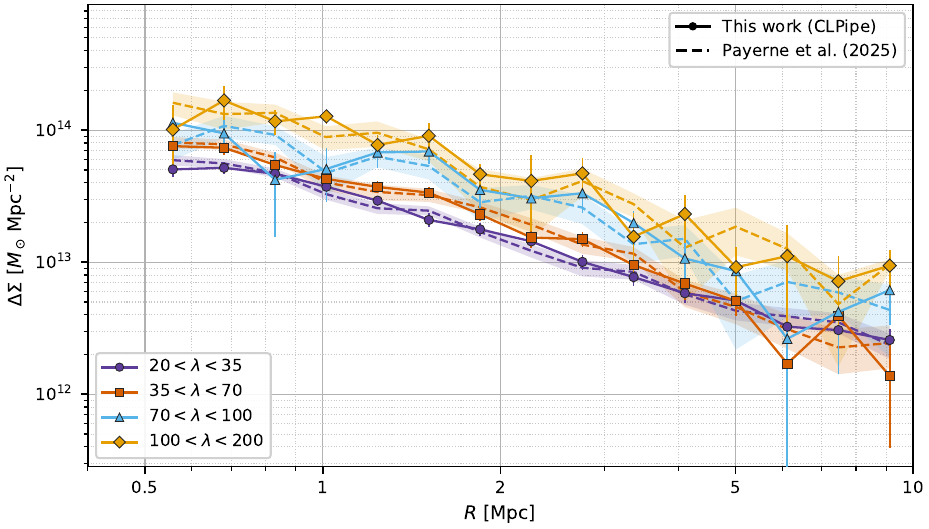}
    \caption{Comparison between \redmapper cosmoDC2 \texttt{TXPipe} lensing data vectors used in this work and \texttt{CLCosmoSim} lensing data vectors from \cite{Payerne2025dc2CL}. Dataset and configurations are described in Sec.~\ref{sec:results}. Filled contours represent \texttt{CLCosmoSim} error bars.}
    \label{fig:profile_comparison}
\end{figure*}

\section{Cross-Validation Against an Independent cosmoDC2-\redmapper Analysis}
\label{app:comparion_payerne2025}

As a validation of the \texttt{TXPipe} procedure, Fig.~\ref{fig:profile_comparison} shows stacked lensing profiles around \redmapper clusters measured by \texttt{TXPipe} in the LSST DESC cosmoDC2 simulated datasets~(see full details in Sec.~\ref{sec:results}). Here we extend the radius binning for comparison purposes. When comparing to \citet{Payerne2025dc2CL} who analyzed the same cluster dataset with the LSST DESC \texttt{CLCosmoSim} code (with different source selection), it shows an overall good agreement.

\begin{figure*}[t]
\centering
\includegraphics[width=0.7\textwidth]{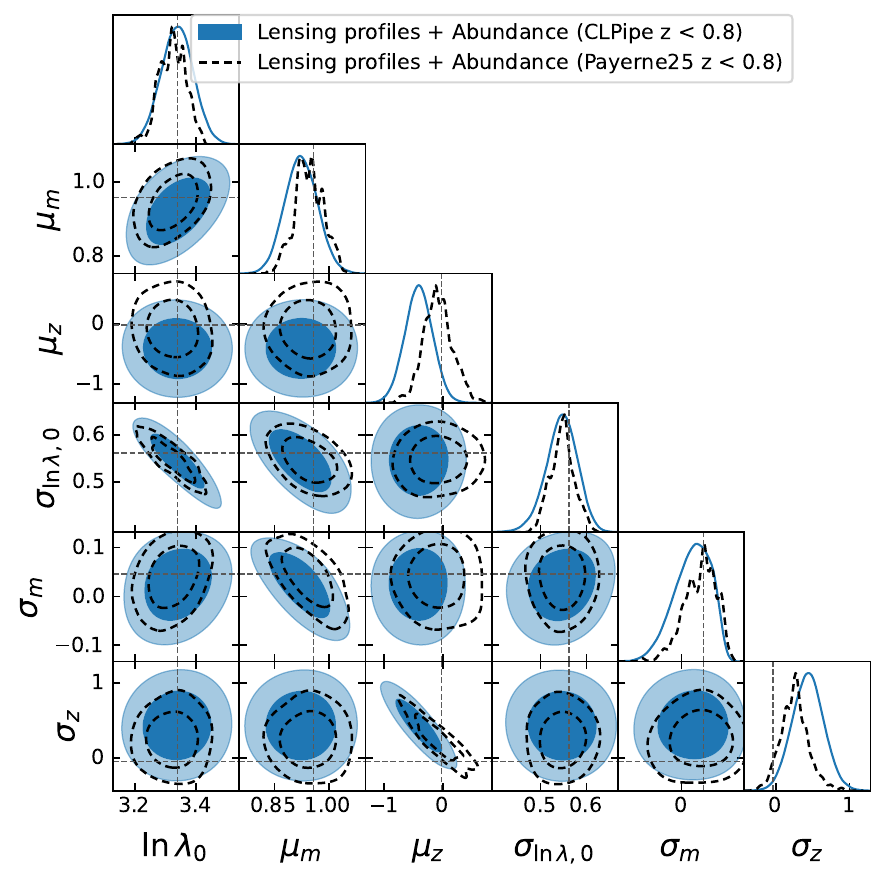}
\caption{Posterior distribution of the mass--richness relation parameters using abundance and weak lensing profiles. \pipeline posteriors are shown in blue and \cite{Payerne2025dc2CL} chains are shown in dashed black lines. Dashed crosses represent the fitted fiducial values from Table~\ref{tab:fiducial}.}
\label{fig:payerne_clp}
\end{figure*}

We perform a fixed-cosmology parameter inference with the same configuration as in Sec.~\ref{sec:results}. Fig.~\ref{fig:payerne_clp} shows the corresponding posterior distribution when combining cluster count and lensing. Our constraints are consistent within $2\sigma$ with the scaling relation fiducial values. The comparison also shows good agreement with the results obtained in \citet{Payerne2025dc2CL}, with a slight difference in $\mu_z$. This can be explained by a difference in the selection function modeling between the two analyses. \citet{Payerne2025dc2CL} model purity as a non-trivial function of redshift and richness, whereas we fix it to unity above our richness threshold ($\lambda \geq 20$; App.~\ref{sec:selfunc}), which could directly impact the contours in the $\mu_z$--$\ln\lambda_0$ plane. Other differences between the two analyses include the source selection used in the data measurement, and the injection of artificial shape noise in our pipeline versus the use of intrinsic galaxy shapes in \citet{Payerne2025dc2CL}.

\end{document}